\documentstyle[epsf,aps]{revtex}
\renewcommand{\thefootnote}{\fnsymbol{footnote}}
\begin{document}
\newcommand{\be}{\begin{eqnarray}}
\newcommand{\dlq}{\lq\lq}
\newcommand{\ee}{\end{eqnarray}}
\newcommand{\ben}{\begin{eqnarray*}}
\newcommand{\een}{\end{eqnarray*}}
\newcommand{\stackeven}[2]{{{}_{\displaystyle{#1}}\atop\displaystyle{#2}}}
\newcommand{\lsim}{\stackeven{<}{\sim}}
\newcommand{\gsim}{\stackeven{>}{\sim}}
\newcommand{\un}[1]{\underline{#1}}
\renewcommand{\baselinestretch}{1.0}
\newcommand{\as}{\alpha_s}
\def\eq#1{{Eq.~(\ref{#1})}}
\def\fig#1{{Fig.~\ref{#1}}}
\begin{flushright}
NT@UW--01--027 \\
INT--PUB--01--25 \\
\end{flushright}
\vspace*{1cm} 
\setcounter{footnote}{1}
\begin{center}
{\Large\bf Inclusive Gluon Production in DIS at High Parton Density}
\\[1cm]
Yuri V.\ Kovchegov $^{1}$ and  Kirill Tuchin $^{2}$ \\ ~~ \\ 
{\it $^1$ Department of Physics, University of Washington, Box 351560} \\ {\it
Seattle, WA 98195 } \\ ~~ \\ 
{\it $^2$ Institute for Nuclear
Theory, University of Washington, Box 351550 } \\ {\it Seattle, WA
98195 } \\ ~~ \\ ~~ \\
\end{center}
\begin{abstract}
We calculate the cross section of a single inclusive gluon production
in deep inelastic scattering at very high energies in the saturation
regime, where the parton densities inside hadrons and nuclei are large
and the evolution of structure functions with energy is nonlinear. The
expression we obtain for the inclusive gluon production cross section
is generated by this nonlinear evolution. We analyze the rapidity
distribution of the produced gluons as well as their transverse
momentum spectrum given by the derived expression for the inclusive
cross section. We propose an ansatz for the multiplicity distribution
of gluons produced in nuclear collisions which includes the effects of
nonlinear evolution in both colliding nuclei.
\end{abstract}
\renewcommand{\thefootnote}{\arabic{footnote}}
\setcounter{footnote}{0}

\section{Introduction}

At very high energies corresponding to very small values of Bjorken
$x$ variable the density of partons in the hadronic and nuclear wave
functions is believed to become very large reaching the {\it
saturation} limit \cite{glrmq,LR,Mue,mv}. In the saturation regime the
growth of partonic structure functions with energy slows down
sufficiently to unitarize the total hadronic cross sections. The
gluonic fields in the saturated hadronic or nuclear wave function are
very strong \cite{yuri,jklw}. A transition to the saturation region
can be characterized by the {\it saturation scale} $Q_s^2 (s)$, which
is related to the typical two dimensional density of the partons'
color charge in the infinite momentum frame of the hadronic or nuclear
wave function \cite{mv,yuri,jklw}. The saturation scale $Q_s^2 (s)$ is
an increasing function of energy $s$ and of the atomic number of the
nucleus A \cite{yurieq,dip,cm,bal,JKLW,lt,braun,agl,lm,lub,gbm}. At
high enough energies or for sufficiently large nuclei the saturation
scale becomes much larger than $\Lambda_{QCD}^2$ allowing for
perturbative description of the scattering process at hand
\cite{mv,musat}. The presence of intrinsic large momentum scale $Q_s$
justifies the use of perturbative QCD expansion even for such a
traditionally non-perturbative observables as total hadronic cross
sections.

Recently there has been a lot of activity devoted to calculating
hadronic and nuclear structure functions in the saturation regime. The
original calculation of quark and gluon distribution functions
including multiple rescatterings without QCD evolution in a large
nucleus was performed in \cite{Mue}. The resulting Glauber-Mueller
formula provided us with expressions for the partonic structure
functions which reach saturation at small $Q^2$. McLerran and
Venugopalan has argued in \cite{mv} that the large density of gluons
in the partonic wave functions at high energy allows one to
approximate the gluon field of a large hadron or nucleus by a
classical solution of the Yang-Mills equations. The resulting gluonic
structure functions has been shown to be equivalent to the
Glauber-Mueller approach \cite{yuri,jklw,KM}. An important problem
which still remained was the inclusion of quantum QCD evolution in
this quasi-classical expression for the structure functions. The
problem was equivalent to resummation of the multiple BFKL pomeron
\cite{BFKL} exchanges.  The evolution equation resumming leading
logarithms of energy ($\as \ln s$) and the multiple pomeron exchanges
was written in \cite{yurieq} using the dipole model of
\cite{dip} and independently in \cite{bal} using the effective high
energy lagrangian approach. The equation was written for the cross
section of a quark-antiquark dipole scattering on a target hadron or
nucleus, which, in turn can yield us the $F_2$ structure function of
the target. The latter can be written as
\be\label{f2}
  F_2 (x, Q^2) = \frac{Q^2}{4 \pi^2 \alpha_{EM}} \int \frac{d^2 r d
  \alpha }{2 \pi} \, \Phi^{\gamma^* \rightarrow q{\bar q}}
  ({\underline r},\alpha) \ d^2 b \ N({\underline r},{\underline b} ,
  Y) ,
\ee
with $\Phi^{\gamma^* \rightarrow q{\bar q}} ({\underline r},\alpha)$
the wave function of a virtual photon in deep inelastic scattering
(DIS) splitting in a $q\bar q$ with transverse separation ${\underline
r}$ and the fraction of photon's longitudinal momentum carried by the
quark $\alpha$. $\Phi^{\gamma^* \rightarrow q{\bar q}} ({\underline
r},\alpha)$ is a very well known function and could be found for
instance in \cite{yurieq,lt,musat}. The quantity $N({\underline
r},{\underline b} , Y)$ has a meaning of a forward scattering
amplitude of a dipole with transverse size $\underline r$ at impact
parameter $\underline b$ with rapidity $Y$ on a target proton or
nucleus normalized in such a way that the total cross section for the
process is given by
\be\label{norm}
\sigma^{q{\bar q} A}_{tot} \, = \, 2 \int d^2 b \, 
 N({\underline r},{\underline b} , Y).
\ee
The evolution equation for $N$ closes only in the large-$N_c$ limit of
QCD \cite{bal,JKLW} and reads \cite{yurieq,dip}
\ben
  N({\underline x}_{01},{\underline b}, Y) = - \gamma ({\underline
  x}_{01},{\underline b}) \, \exp \left[ - \frac{4 \alpha C_F}{\pi} \ln
  \left( \frac{x_{01}}{\rho} \right) Y \right] + \frac{\alpha
  C_F}{\pi^2} \int_0^Y d y \, \exp \left[ - \frac{4 \alpha C_F}{\pi}
  \ln \left( \frac{x_{01}}{\rho} \right) (Y - y) \right]
\een
\be\label{eqN}
\times \int_\rho d^2 x_2 \frac{x_{01}^2}{x_{02}^2 x_{12}^2} \, [ 2
  \, N({\underline x}_{02},{\underline b} + \frac{1}{2} {\underline
  x}_{12}, y) - N({\underline x}_{02},{\underline b} + \frac{1}{2}
  {\underline x}_{12}, y) \, N({\underline x}_{12},{\underline b} +
  \frac{1}{2} {\underline x}_{02}, y) ] ,
\ee
with the initial condition set by $\gamma ({\underline
x}_{01},{\underline b})$, which is the propagator of a dipole of size
${\underline x}_{01}$ at the impact parameter ${\underline b}$ through
the target nucleus or hadron. $\gamma$ was taken to be of
Mueller-Glauber form in \cite{yurieq}:
\begin{equation}\label{gla}
   \gamma ({\underline x}_{01},{\underline b}_0) = e^{ -
   {\underline x}_{01}^2 Q_{0s}^{quark 2} / 4} - 1,
\end{equation}
where for a spherical nucleus \cite{Mue,KM,yuincl}
\be\label{xqs}
{\underline x}_{01}^2 Q_{0s}^{quark 2} \, = \, {\underline x}_{01}^2 \ \frac{4
\pi^2 \as \sqrt{R^2 - b^2}}{N_c} \, \rho \, 
xG (x, 1/{\underline x}_{01}^2),
\ee
with $\rho = A / [(4/3) \pi R^3]$ the density of the atomic number
$A$. The gluon distribution in \eq{xqs} should be taken at the
two-gluon order
\be\label{xg}
xG (x, 1/{\underline x}^2) \ = \ \frac{\as C_F}{\pi} \, \ln
\frac{1}{{\underline x}^2 \mu^2},
\ee
with $\mu$ some infrared cutoff. The scale $Q_{0s}^{quark 2}$ has the
meaning of the quasi-classical quark saturation scale generated by
multiple rescatterings prior to the inclusion of
evolution. Eqs. (\ref{f2}) and (\ref{eqN}) provide us with the $F_2$
structure function and the total cross section of DIS on a nucleus
including all multiple BFKL pomeron exchanges (fan diagrams). In spite
of several attempts to solve \eq{eqN} analytically, which provided us
with a well-understood high and low energy asymptotics for $N$
\cite{yurieq,lt,JKLW}, the exact analytical solution is still to be
found. There exist several numerical solutions to \eq{eqN}
demonstrating that at very high energies the amplitude $N$ goes to an
independent of energy constant ($N \rightarrow 1$) thus unitarizing
the total DIS cross sections \cite{lt,braun,lub,gbm}. The numerical
analyses also show that \eq{eqN} does generate a momentum scale $Q_s$
which rapidly increases with energy. This scale justified the small
coupling $\as$ expansion and helps avoid the problem of infrared
instability of BFKL equation. An effort to calculate the NLO
correction to \eq{eqN} is currently under way \cite{bb}.

Several other observables could be calculated in the framework of the
saturation approach to hadronic and nuclear collisions. Of the
exclusive observables diffractive (or, more precisely, elastic) cross
section has been calculated in the quasi-classical limit in
\cite{diffqc}. The evolution equation including multiple pomeron
exchanges has been recently written for the cross section of single
diffractive dissociation in \cite{kl}.

In this paper we are interested in inclusive particle production cross
sections. In the saturation framework these cross sections have been
extensively studied at the classical level in DIS, proton--nucleus
(pA) and nucleus--nucleus (AA) collisions. The inclusive gluon
production cross section for pA and AA in the weak classical field
limit (lowest order in perturbation theory without multiple
rescatterings) has been calculated in \cite{claa}, reproducing the
result of Gunion and Bertsch \cite{gb}. In the strong field limit
including all multiple rescatterings but no QCD evolution the gluon
production cross section was calculated for pA in \cite{KM}. This
result has been recently reproduced in \cite{kop,md,kw}. The inclusive
gluon production cross section for DIS was calculated in
\cite{yuincl}, with the resulting expression being slightly different 
from a straightforward generalization of the pA result of
\cite{KM}. Finally an important problem for heavy ion physics is the
calculation of the inclusive gluon production cross section in AA,
which would provide us with the initial conditions for the possible
formation of quark-gluon plasma. Numerical estimates of the related
gluon multiplicities were performed in \cite{kv}, while an analytical
ansatz has been proposed in \cite{yuaa}.

The problem of inclusion of nonlinear evolution in the inclusive cross
sections received much less attention in the literature. The case of
heavy flavor production in DIS with nonlinear evolution has been
solved in \cite{ab}. The inclusive gluon production in DIS has been
studied in \cite{glrjet,brjet} using the AGK cutting rules \cite{agk}
and $k_T$-factorization approach.

In this paper we calculate the single inclusive gluon production cross
section in DIS including the effects of multiple rescatterings and
nonlinear evolution of \eq{eqN}. We begin in Sect. II by reviewing the
derivation of the formula for the single inclusive gluon production
cross section in the quasi-classical approximation given in
\cite{yuincl}. In the quasi-classical approximation the quantum evolution 
is not included since one is interested in resummation of multiple
rescatterings \cite{yuri,KM}. Each multiple rescattering of the
produced gluon on a nucleon in the nucleus brings in a factor of
$\as^2 A^{1/3}$ and the quasi-classical approximation can be defined
as resumming powers of this parameter \cite{yuri,musat,KM,yuaa}. The
result for gluon production cross section is shown in \eq{qcincl}.

We continue in Sect. III by including the effects of nonlinear dipole
evolution from \eq{eqN} in the expression for the cross section. The
philosophy of our approach is similar to \cite{yurieq}. We first
construct a classical Glauber-Mueller--type expression for the
inclusive cross section and then use it as our starting point for
including dipole evolution. We are working in the rest frame of the
target, which allows to consider the quantum evolution in energy as
happening only in the wave function of the incoming $q{\bar q}$ pair
\cite{yurieq,bal}. In Sect. IIIA we analyze the evolution preceding 
the emission of the gluon that we measure in the final state. This
evolution corresponds to emission of gluons with larger (harder) light
cone component of momentum than the one carried by the gluon that we
trigger. We show that this early evolution can only be linear (single
pomeron exchange) leading to creation of the dipole in which our
measured gluon is emitted. This conclusion is in agreement with the
prediction of AGK cutting rules for inclusive cross section
\cite{glrjet,brjet,agk,gein}. In Sect. IIIB we proceed by analyzing the 
emissions of the gluons which are softer than the measured one. We
demonstrate that the effect of these later emissions can be
incorporated in the inclusive cross section by replacing the
Glauber-Mueller exponents in \eq{qcincl} with the evolved forward
amplitudes of a gluon (adjoint) dipole scattering on the target. The
final expression for inclusive gluon production cross section in DIS
is given by \eq{incl} in Sect. IIIC which is the main result of this
paper.

We begin analyzing the cross section of \eq{incl} in Sect. IV by
observing that in the case of a very large nucleus corresponding to
zero momentum transfer in each of the exchanged pomerons it can be
rewritten in a factorized form as a convolution of two functions with
Lipatov's effective vertex inserted in the middle (see \eq{fact}). In
this form it almost agrees with the expression derived by Braun in
\cite{brjet} using the AGK rules and pomeron fan diagram approach (see
Eq. (10) in \cite{brjet}). To obtain the expression given in
\cite{brjet} from our \eq{fact} one has to replace in it the evolved 
forward amplitude of an adjoint dipole on the target nucleus $N_G$ by
the similar amplitude for the fundamental dipole $N$. While the
difference seems minor in the weak field limit given by the linear
evolution equations, it becomes much more profound in the transition
to the saturation region where the quantities $N_G$ and $N$ obey
different evolution equations while approaching the same high energy
asymptotics. In Sect. V we study the transverse momentum spectrum and
energy dependence of the obtained cross section. We observe that while
at very large $k_\perp$ the cross section exhibits the usual
$1/k_\perp^4$ behavior it softens to $1/k_\perp^2$ in the small
$k_\perp \sim Q_s$ region similarly to the quasi-classical cross
sections of \cite{KM,yuincl}. The rapidity distribution of the
produced gluons may have a maximum which position is determined by the
values of the produced gluon's momentum $k_\perp$ and photon's
virtuality $Q$ (see \fig{ydep}).

We conclude in Sect. VI by discussing the general principles of
inclusion of nonlinear evolution in the cross sections for various
inclusive processes. We present an ansatz for the multiplicity
distribution of the produced gluons in nucleus--nucleus collisions
(AA) including the effects of nonlinear evolution in both colliding
nuclei.

\section{Inclusive Cross Section in the Quasi--Classical Approximation}

\begin{figure}
\begin{center}
\epsfxsize=13cm
\leavevmode
\hbox{ \epsffile{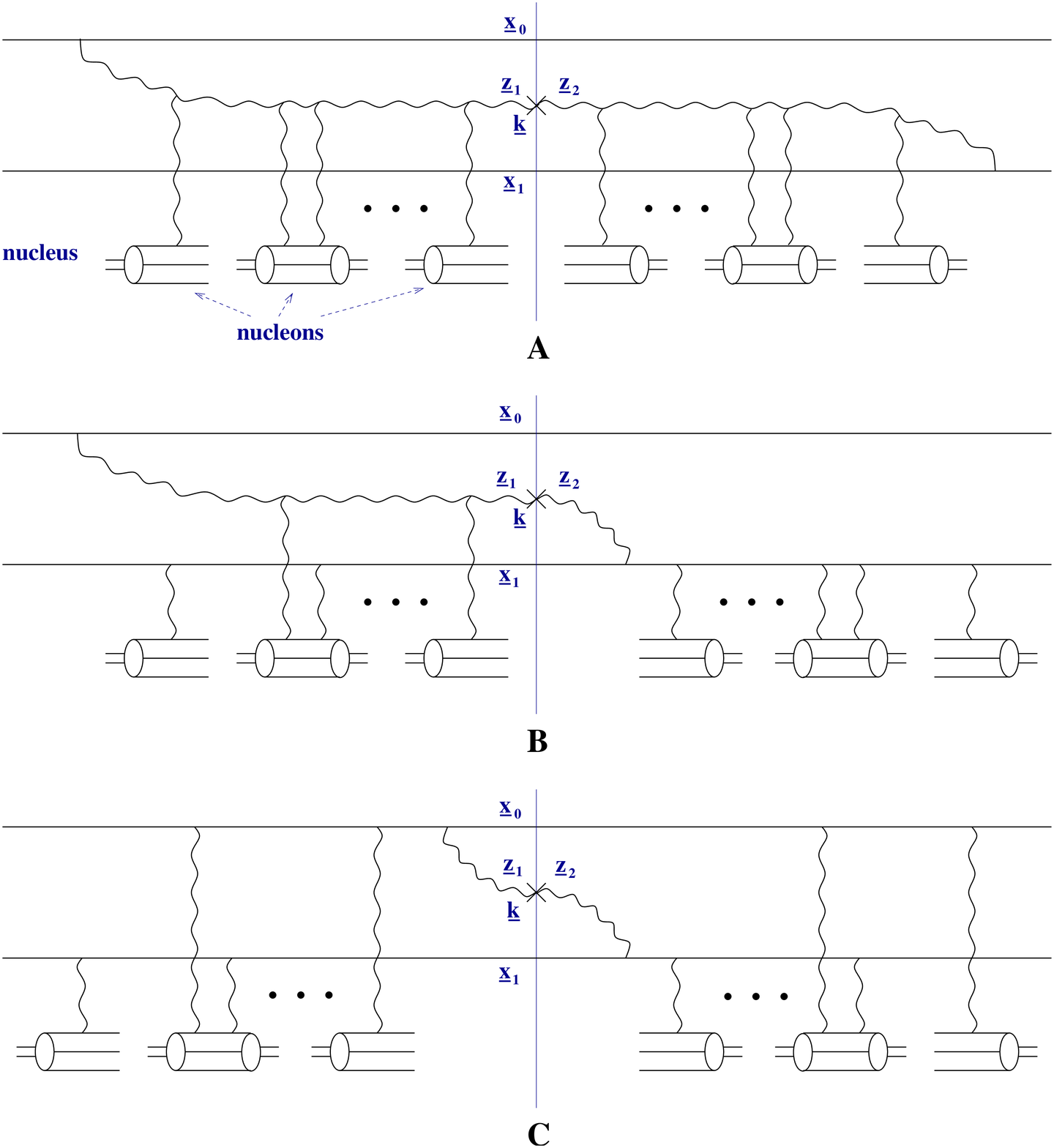}}
\end{center}
\caption{Gluon production in DIS in the quasi-classical approximation. 
The produced gluon may be emitted either off the quark or off the
antiquark lines both in the amplitude and in the complex conjugate
amplitude. Only one connection is shown. }
\label{qcdis}
\end{figure}

In this Section we will review the derivation of the single inclusive
gluon production cross section in DIS from \cite{yuincl} in the
quasi-classical approximation developed in \cite{KM}. The gluon
production cross section in DIS can be rewritten as
\be\label{disincl}
\frac{d \sigma^{\gamma^* A \rightarrow q{\bar q}GX}_{incl}}{d^2 k \, dy} \, 
= \, \frac{1}{2 \pi^2} \, \int \, d^2 x_{01} \, d \alpha \, \Phi^{\gamma^*
\rightarrow q\bar q} ({\underline x}_{01}, \alpha) \, \frac{d {\hat
\sigma}^{q{\bar q}A}_{incl}}{d^2 k \, dy}({\underline x}_{01}),
\ee
where $d {\hat \sigma}^{q{\bar q}A}_{incl}/d^2 k \, dy ({\underline
x}_{01})$ is the gluon production cross section for the scattering of
a dipole of transverse size ${\underline x}_{01} = {\un x}_0 - {\un
x}_1$ on the target. We want to calculate this observable including
all multiple rescatterings of the $q\bar q$ pair and the produced
gluon on the nucleons in the target nucleus. In the quasi-classical
approximation each rescattering happens via one or two gluon exchanges
with the nucleon \cite{KM}. We assume that the initial quark-antiquark
pair is moving in the ``+'' light cone direction. The diagrams
contributing to the gluon production cross section in the $A_+ = 0$
light cone gauge are shown in \fig{qcdis}.

Similar to \cite{KM,yuincl,kop,md,kw,yuaa} the gluon emission can
happen via two possible scenarios: the incoming dipole may have the
gluon fluctuation in its wave function by the time it hits the target
or the gluon may be emitted after the dipole interacts with the
target. Even with all the multiple rescatterings the interaction with
the target is instantaneous compared to the typical emission time of
the gluon \cite{KM}. We thus may denote the interaction time $\tau
\equiv x_+ = 0$. If the gluon emission time in the amplitude is 
$\tau_1$ and in the complex conjugate amplitude is $\tau_2$ than the
following classification of diagrams in \fig{qcdis} is possible. The
graph in \fig{qcdis}A corresponds to the case when $\tau_1 < 0$ and
$\tau_2 < 0$, while the diagram in \fig{qcdis}B reflects the $\tau_1 <
0$, $\tau_2 > 0$ case. A ``mirror image'' diagram should be added to
\fig{qcdis}B describing the $\tau_1 > 0$, $\tau_2 < 0$ case. The late 
emission $\tau_1 > 0$, $\tau_2 > 0$ scenario is represented in
\fig{qcdis}C. The produced gluon line can start off either quark or 
antiquark lines both in the amplitude and in the complex conjugate
amplitude in \fig{qcdis}. We are going to sum over all possible
emissions while only one case is shown in \fig{qcdis}.

To calculate the diagrams in \fig{qcdis} we will be working in
transverse coordinate space with the intent to perform a Fourier
transform into momentum space in the end. Thus the produced gluon has
different transverse coordinates in the amplitude and in the complex
conjugate amplitude (${\underline z}_1$ and ${\underline z}_2$). The
dipole consists of the quark at ${\underline x}_0$ and an antiquark at
${\underline x}_1$. Due to real-virtual cancellations only the
diagrams where the nucleons interact with the produced gluon survive
in \fig{qcdis}A \cite{KM}. The square of the gluon's propagator could
be easily calculated to give \cite{KM}
\be\label{qc1}
e^{- ({\underline z}_1- {\underline z}_2)^2 Q_{0s}^2 /4}
\ee
with
\be\label{gs}
{\underline x}^2 Q_{0s}^2 \, = \, {\underline x}^2 \ \frac{8 \pi^2 \as
N_c \sqrt{R^2 - b^2}}{N^2_c - 1} \, \rho \, xG (x, 1/|{\underline
x}|^2).
\ee
The scale $Q_{0s}^2$ has the meaning of the gluon saturation scale in
the quasi-classical (no evolution) approximation and is different from
the quark saturation scale of \eq{xqs} by the Casimir operator. In
\fig{qcdis}B only the interactions with the gluon line and the
antiquark are allowed. Performing the calculation along the lines of
Appendix A in \cite{KM} and adding the ``mirror'' contribution one
obtains
\be\label{qc2}
- e^{- ({\underline z}_1- {\underline x}_1)^2 Q_{0s}^2 /4} - e^{-
({\underline z}_2- {\underline x}_0)^2 Q_{0s}^2 /4}.
\ee
The minus sign in \eq{qc2} is due to the fact that the gluon is
emitted after the interaction on one side of the cut in
\fig{qcdis}B. Finally the interactions with the nucleons shown in 
\fig{qcdis}C would have canceled if we were considering proton-nucleus 
(or, more precisely, quark-nucleus) collisions as it was done in
\cite{KM}. However, in the DIS case at hand the quark and antiquark 
originate from a virtual photon and thus have to be in the color
singlet initial state on both sides of the cut. This condition was not
imposed in pA \cite{KM,kw} where the color of the interacting quark
was assumed to be randomized by the non-perturbative ``intrinsic''
quarks and gluons in the proton's wave function
\cite{brod}. Therefore, unlike the pA case moving an exchanged gluon 
line across the cut in DIS would modify the color factor of the
diagram and thus real-virtual cancellation would not happen. A
calculation of the interactions gives 
\be\label{qc3}
e^{- ({\underline x}_0- {\underline x}_1)^2 Q_{0s}^2 /4}
\ee
for the diagram in \fig{qcdis}C.

To obtain the final answer we now have to combine the terms from
Eqs. (\ref{qc1}), (\ref{qc2}) and (\ref{qc3}), multiply by the
amplitude of gluon's emission in the original dipole while keeping in
mind that the amplitudes in \fig{qcdis} are slightly different for
different connections of the gluon to the $q\bar q$ pair. We then
should Fourier transform the expression into momentum space. The final
result for the single inclusive gluon production cross section of
dipole-nucleus scattering in the quasi--classical approximation reads
\ben
\frac{d {\hat \sigma}^{q{\bar q}A}_{incl}}{d^2 k \, dy}({\underline x}_{01}) 
\, = \,  \frac{\as C_F}{\pi^2} \, \frac{1}{(2 \pi)^2} \, \int \, d^2 b \, 
d^2 z_1 \, d^2 z_2 \, e^{- i {\underline k} \cdot ({\underline z}_1 -
{\underline z}_2)} \, \sum_{i,j=0}^1 (-1)^{i+j}
\frac{{\underline z}_1- {\underline x}_i}{|{\underline z}_1-
{\underline x}_i|^2} \cdot
\frac{{\underline z}_2- {\underline x}_j}{|{\underline z}_2- {\underline
x}_j|^2} 
\een
\be\label{qcincl}
\times \left( e^{- ({\underline x}_i- {\underline
x}_j)^2 Q_{0s}^2 /4} - e^{- ({\underline z}_1- {\underline x}_j)^2
Q_{0s}^2 /4} - e^{- ({\underline z}_2- {\underline x}_i)^2 Q_{0s}^2
/4} + e^{- ({\underline z}_1- {\underline z}_2)^2 Q_{0s}^2 /4} \right)
\ee
where ${\underline b} = ({\underline x}_0 + {\underline x}_1)/2$ is
the dipole's impact parameter.  Together Eqs. (\ref{disincl}) and
(\ref{qcincl}) give us the inclusive gluon production cross section in
DIS in the quasi--classical approximation as derived in \cite{yuincl}.

\section{Inclusion of Evolution Effects}

We are now going to include quantum evolution into \eq{qcincl}. We
begin in Sect. IIIA by discussing the evolution preceding the emission
of the measured gluon, and continue in Sect. IIIB by analyzing the
subsequent evolution. The final expression is derived in Sect. IIIC.

\subsection{Emission of Harder Gluons}

\begin{figure}
\begin{center}
\epsfxsize=15cm
\leavevmode
\hbox{ \epsffile{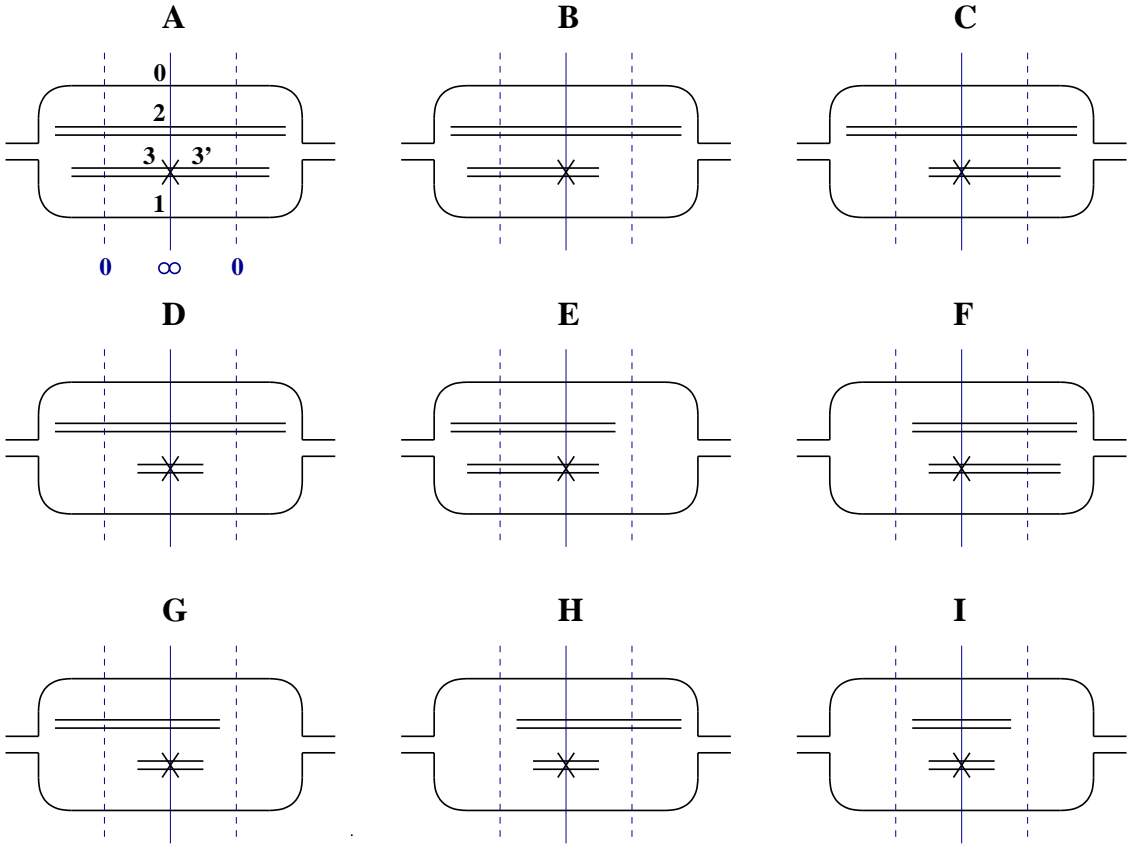}}
\end{center}
\caption{Emission of a harder gluon in the dipole evolution. Gluons are 
denoted by double lines in the large $N_c$ limit. The produced gluon
is marked by a cross.}
\label{eem}
\end{figure}

Let us explore how the emission of gluons with the light cone ``plus''
component of the momentum much larger than that of the measured gluon
modify the inclusive cross section. For simplicity we first consider
an emission of a single extra gluon. The diagrams relevant for real
emission of this extra gluon are shown in \fig{eem}. Since we are
interested in the large-$N_c$ dipole evolution all gluons are
represented by double quark lines. Notations are explained in
\fig{eem}A. The incoming original dipole of transverse size
${\underline x}_{01} = {\underline x}_0 - {\underline x}_1$ emits a
(harder) gluon with transverse coordinate ${\underline x}_2$. Then the
measured gluon ($\# 3$) is emitted in one of the color dipoles formed
by the emission of gluon $\# 2$. Since we are interested in keeping
this gluon's transverse momentum fixed in the final state its
transverse coordinates are different on each side of the cut
(${\underline x}_3$ and ${\underline x}_{3'}$).  Emission in the lower
dipole only is shown in \fig{eem}. Emission in the upper dipole is
completely analogous and could be obtained from \fig{eem} by switching
gluons $2$ and $3$.

The gluon lines in \fig{eem} can connect to either quark or antiquark
lines in the dipoles off which they are emitted both in the amplitude
and in the complex conjugate amplitude. This is shown in \fig{eem} by
not connecting the gluon lines to any of the quark line
specifically. For instance the gluon $\# 2$ can be emitted off the
quark line $0$ and off the antiquark line $1$, which is demonstrated
by drawing gluon $\# 2$ directly between those quark lines. This
notation is the same as the one used in \cite{cm,kl}. The interactions
with the target are not shown explicitly in \fig{eem}. Instead we mark
with a dashed vertical line in \fig{eem} the moment in light cone time
$\tau = 0$ when the multiple rescatterings of \fig{qcdis} occur (see
\fig{eem}A). Of course the interactions may happen both in the amplitude and 
in the complex conjugate amplitude. The solid vertical line in
\fig{eem} denotes the final state corresponding to $\tau = \infty$.

\begin{figure}[t]
\begin{center}
\epsfxsize=12cm
\leavevmode
\hbox{ \epsffile{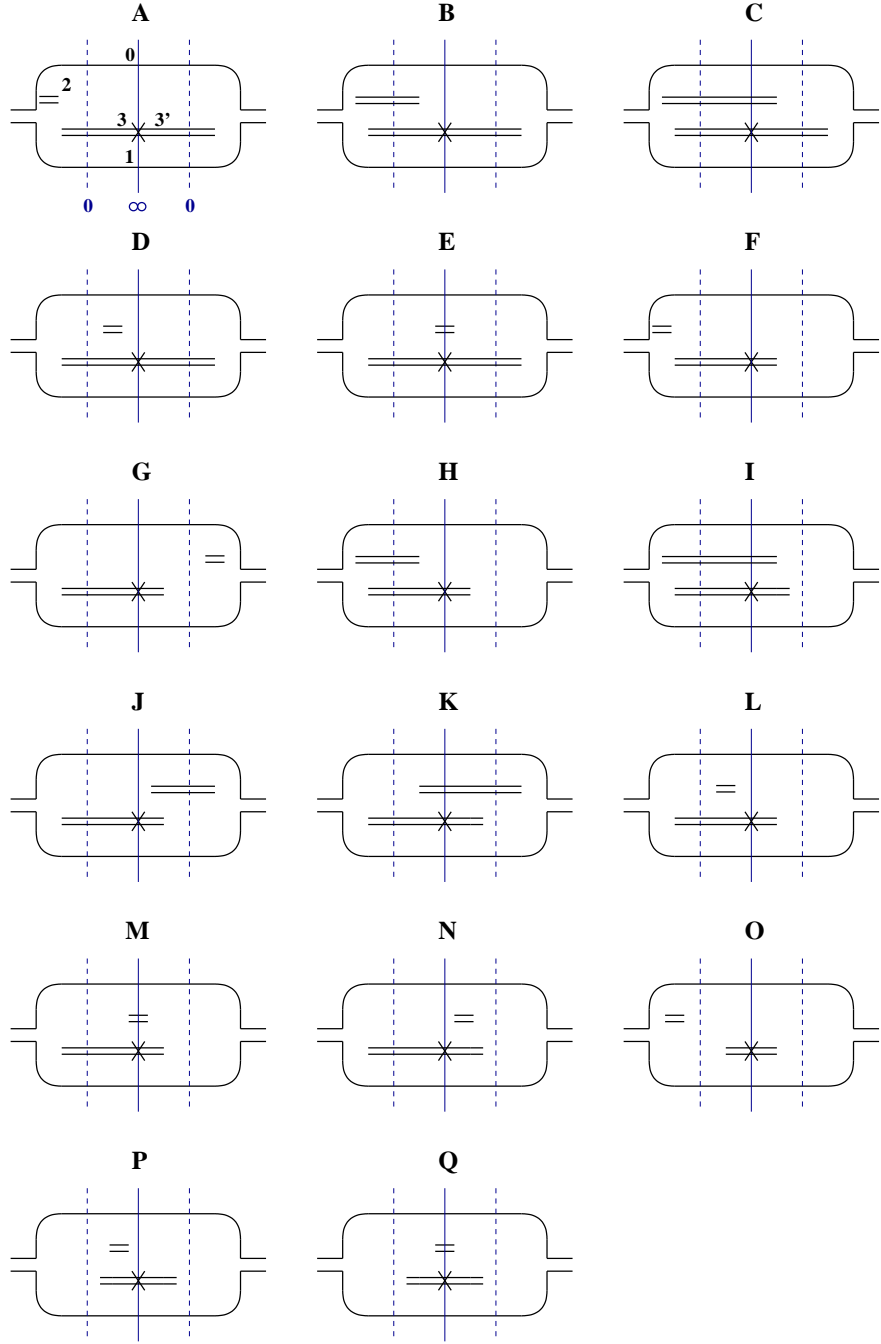}}
\end{center}
\caption{Another set of graphs contributing to emission of a harder gluon 
in the dipole evolution.}
\label{vem}
\end{figure}

The measured gluon $\# 3$ is much softer than the gluon $\# 2$,
$k_{3+} \ll k_{2+}$, while their transverse momenta are not
ordered. Thus the lifetime of the gluon $\# 2$ is much longer than
that of gluon $\# 3$. Therefore it seems natural that in all the
diagrams considered in \fig{eem} the gluon $\# 2$ is emitted before
gluon $\# 3$ on both sides of the cut. However the graphs in \fig{eem}
do not cover all the diagrams which need to be considered. There is
another set of diagrams, predominantly with the virtual emission of
gluon $\# 2$, some of which are shown in \fig{vem}, that could be
important for gluon production at this order. We are going to first
analyze the diagrams in \fig{eem}, which would allow us to understand
which ones of the diagrams omitted in \fig{eem} should be considered.

The transverse momentum $\underline k$ and rapidity $y$ of the
measured gluon $\# 3$ are kept fixed, while the transverse momentum
and rapidity of the other emitted gluon are integrated over. In the
spirit of the leading logarithmic evolution integration over the
rapidity of the gluon $\# 2$ is supposed to give us the factor of $\ln
1/x$ with $x$ the Bjorken $x$ variable, or, equivalently, a factor of
total rapidity interval $Y$. This factor would make up for the
suppression by the power of coupling constant $\as$ that we have
introduced by emitting that gluon. Now we are going to demonstrate
that this enhancement does not happen in
Figs. \ref{eem}E--\ref{eem}I. That implies that the diagrams in
Figs. \ref{eem}E--\ref{eem}I do not give a leading logarithmic
contribution and could be neglected as subleading.

\begin{figure}
\begin{center}
\epsfxsize=5cm
\leavevmode
\hbox{ \epsffile{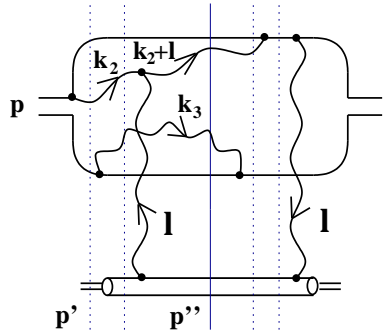}}
\end{center}
\caption{One of the diagrams contributing to the class of graphs represented 
by \fig{eem}E.}
\label{rem}
\end{figure}

Let us calculate the diagram in \fig{rem}, which is one of the graphs
represented by \fig{eem}E. The diagrams in \fig{eem} are understood to
be in the light cone perturbation theory (LCPT) \cite{brod1}, which
has been used in the original construction of the QCD dipole model
\cite{dip}. Therefore we will use the diagrammatic rules of LCPT in 
evaluating the graph \fig{rem}. We are working in the frame where the
quarks in the original dipole have large ``$+$'' components of their
momenta while the nucleons in the nucleus have large ``$-$'' momentum
components. Though only one rescattering is depicted in
\fig{rem} our conclusions will be easy to generalize to include
multiple Glauber rescatterings of \fig{qcdis}. The intermediate states
are denoted by vertical dotted lines in \fig{rem}. In estimating the
energy denominators of the intermediate states at $\tau > 0$ one
should remember that the ``$-$'' component of the nucleon's momentum
also changes. According to the rules of LCPT the ``$-$'' component of
the final state should be equal to the ``$-$'' component of the
initial (incoming) state \cite{brod1}. Thus one requires that for the
diagram in \fig{rem} the following condition should be satisfied
\be\label{cons}
p''_- + \frac{({\un k}_2 + {\un l})^2}{2 k_{2+}} + \frac{{\un
k}_3^2}{2 k_{3+}} \, = \, p'_-,
\ee
where we have used that for exchanged Coulomb gluons $l_- = 0$
\cite{yuri,KM}. Using \eq{cons} in evaluating, for instance, the energy 
denominator of the rightmost intermediate state in \fig{rem} yields
\cite{brod1}
\be\label{smpl}
\frac{1}{p''_- - p'_-} \, = \, \frac{1}{- \frac{({\un k}_2 + 
{\un l})^2}{2 k_{2+}} - \frac{{\un k}_3^2}{2 k_{3+}}} \, \approx \, 
\frac{1}{- \frac{{\un k}_3^2}{2 k_{3+}}}.
\ee
We have used the fact that $k_{3+} \ll k_{2+}$ in the last
approximation in \eq{smpl}. The rule for calculating the diagrams in
\fig{eem} could be formulated as follows: the energy denominators for 
intermediate states with $\tau > 0$ should consist of the sum of the
``$-$'' momenta components of all the intermediate gluons in the state
minus the ``$-$'' momenta components of all the gluons in the final
state. (Note that according to the rules of LCPT the ``$-$'' momenta
components are not conserved at the vertices and all intermediate
lines are on mass shell \cite{brod1}.)

Using the outlined strategy one can show that the contribution of the
diagram in \fig{rem} is proportional to
\be\label{smcalc}
\int_{k_{3+}}^{p_+} d k_{2+} \sum_{\lambda_2 , \lambda_3} 
\, \frac{{\un \epsilon}_2^{\lambda_2} \cdot {\un k}_2 \, {\un \epsilon}_2^{\lambda_2}
\cdot ({\un k}_2 + {\un l}) \, ({\un \epsilon}_3^{\lambda_3} \cdot {\un k}_3)^2}
{({\un k}_3^2)^3 \, {\un k}_2^2 \, k_{2+}^2} \, = \,
\int_{k_{3+}}^{p_+} d k_{2+} \frac{{\un k}_2 \cdot ({\un k}_2 + {\un
l})} {({\un k}_3^2)^2 \, {\un k}_2^2 \, k_{2+}^2} \, \approx \,
\frac{{\un k}_2 \cdot ({\un k}_2 + {\un l})} {({\un k}_3^2)^2 \, {\un
k}_2^2 \, k_{3+}} \, \sim \, \frac{1}{k_{3+}}
\ee
with $p_+$ a large momentum of one of the quarks in the original
dipole ($k_{3+} \ll p_+$). As can be seen from \eq{smcalc} the diagram
in \fig{rem} does not contain any logarithms of energy. It should be
compared to the contribution of, for instance, diagram in
\fig{eem}A, which is proportional to
\be\label{alog}
A \, \sim \, \int_{k_{3+}}^{p_+} d k_{2+} \frac{1}{k_{2+} \, k_{3+}}
\, = \, \frac{1}{k_{3+}} \, \ln \frac{p_+}{k_{3+}}.
\ee
This diagram is enhanced by extra logarithm of energy (or,
equivalently, an extra factor of rapidity) and should be included in
our leading logarithmic analysis. In the same approximation the
diagram in \fig{rem} is subleading since it does not have a logarithm
of energy in it. Similar calculations can be carried out for all the
diagrams in \fig{eem}E showing that they are subleading and should be
neglected. Finally the same conclusion will be reached if one analyzes
all diagrams in Figs. \ref{eem}E--\ref{eem}I : they do not bring in an
extra logarithm of energy and therefore must be neglected.

Here we would like to remind the reader about the approximation we are
using. It is the same approximation as outlined in \cite{yurieq} for
the calculation of total DIS cross section. The (nonlinear) quantum
evolution resums the powers of $\as N_c Y$ and is a function of this
parameter \cite{dip}. ($Y$ is the rapidity variable.) At the same time
we are interested in resumming multiple rescatterings, which also
bring in extra powers of $\as$ \cite{yuri,KM,yuaa}. The multiple
rescatterings are limited to two gluon exchanges with each
nucleon. Thus the actual parameter which is being resummed by multiple
rescatterings is $\as^2 A^{1/3}$ \cite{yuri,KM,yuaa}. Therefore the
combination of leading logarithmic large-$N_c$ evolution and
Glauber-Mueller multiple rescatterings resums all powers of both $\as
N_c Y$ and $\as^2 A^{1/3}$ in the cross sections. From this point of
view the diagrams in Figs. \ref{eem}E--\ref{eem}I are suppressed by an
extra power of $\as N_c$ not enhanced by an extra power of $Y$.

We have proven that of all graphs in \fig{eem} only diagrams in
Figs. \ref{eem}A--\ref{eem}D contribute. Based on this information one
may conjecture that the evolution preceding the emission of the
measured gluon consists of gluons emitted before interaction both in
the amplitude and in the complex conjugate amplitude. To complete the
proof of this statement we have to consider a whole class of diagrams
where the line of gluon $\# 2$ is shorter or of the same length as the
line of gluon $\# 3$. These diagrams include virtual diagrams where
gluon $\# 2$ is not present in the final state. Many of these diagrams
are also subleading, similar to Figs. \ref{eem}E--\ref{eem}I. This can
be seen by performing the analysis outlined above and using the rule
presented in the Appendix of \cite{cm} for calculation of virtual
contributions. At the end we are left with the diagrams shown in
\fig{vem}, each of which gives a leading logarithmic contribution. Not all of the 
diagrams in \fig{vem} are symmetric with respect to horizontal
(left-right) mirror reflections. Therefore one should add mirror
images to diagrams A--D, F--P which will be denoted by primes
(e.g. A', B', etc.). We note that due to momentum conservation the
gluon $\# 2$ can be emitted only off the quark lines $0$ and $1$ in
the original dipole. Therefore moving part or all of gluon $\# 2$
across the cut does not change the transverse coordinate (and,
therefore, transverse momentum) structure of the diagrams. Using the
cancellation of final state interactions demonstrated in \cite{cm} we
observe that in
\fig{vem}
\begin{mathletters}
\be
B + C \, = \, B' + C' \, = \, 0
\ee
\be
D + E + D' = \, 0
\ee
\be
H + I \, = \, H' + I' \, = \, 0
\ee
\be
J + K \, = \, J' + K' \, = \, 0
\ee
\be
L + M + N \, = \, L' + M' + N' \, = \, 0
\ee
\be
P + Q + P' \, = \, 0.
\ee
\end{mathletters}
We are left only with the diagrams A, F, G, O and their mirror images
A', F', G', O' in \fig{vem}.

Combining the results of the analyses of the diagrams in
Figs. \ref{eem} and \ref{vem} we conclude that only the diagrams where
gluon $\# 2$ is either emitted or both emitted and absorbed before the
interaction at $\tau < 0$ survive. The non-vanishing diagrams are
graphs A--D in \fig{eem} and A, F, G, O, A', F', G', O' in
\fig{vem}. One can easily check that these diagrams add up to give us
one rung of the linear dipole (or, equivalently, BFKL) evolution. The
diagrams A--D in \fig{eem} provide us with the real part of the dipole
kernel, splitting the original dipole $01$ in two, in one of which we
continue the evolution by emitting the measured gluon $\# 3$. Diagrams
A, F, G, O, A', F', G', O' of \fig{vem} yield us with virtual
corrections to the same process.

Now we are in a position to generalize our conclusions to the case of
many harder gluons in the virtual photon's wave function. The
evolution preceding the emission of the measured gluon is the usual
early-time dipole evolution leading to creation at early times (before
the interaction, $\tau < 0$) of the dipole in which the measured gluon
is emitted. The effect of this evolution is to modify the probability
of finding this dipole in the virtual photon's wave function.

This preceding evolution can only be linear. This can be understood by
analyzing diagrams A--D in \fig{eem}. Real emissions in the dipole
model may lead to branching of one dipole evolution into two
simultaneous evolutions, being thus equivalent to triple pomeron
vertices in the traditional language \cite{yurieq,dip,braun,robi}. For
instance, in Figs. \ref{eem}A--\ref{eem}D there could in principle be
some subsequent evolution in the dipole $02$ leading to creation of
many dipoles which would interact with the target. The evolution may
also lead to creation of dipoles at late times ($\tau > 0$). However
all these emissions and interactions would not modify the momentum of
the gluon that we measure, since they happen in a different dipole
isolated from ours in the large-$N_c$ limit. Therefore one can show
that all these interactions would cancel via real--virtual
cancellations: first one can show the cancellation of Coulomb gluon
exchanges similarly to some of the cancellations in Sect. II and then
the cancellation of evolution in dipole $02$ follows due to
probability conservation (see \cite{dip,cm}). This can be done for any
graph where the evolution branches into two with the emitted gluon
being produced by one of the subsequent evolutions. We therefore
conclude that the early evolution is {\it linear}. This conclusion is
similar to what one would obtain applying AGK cutting rules to the
process \cite{glrjet,brjet,agk,gein}. We will return to this
similarity in the next Section.

To include the preceding evolution into the gluon production cross
section of \eq{disincl} one should thus substitute
\be\label{levol}
\int \, d^2 x_{01} \, d \alpha \, \Phi^{\gamma^*
\rightarrow q\bar q} ({\underline x}_{01}, \alpha) \, \rightarrow \, \int d^2 r 
 \, d \alpha \, \Phi^{\gamma^* \rightarrow q{\bar q}} ({\underline
 r},\alpha) \ d^2 B \ n ({\underline r}, {\underline x}_{01},
 {\underline B} - {\underline b}, Y - y) \, \frac{d^2 x_{01}}{2 \pi
 x^2_{01}} \,
\ee
where $n ({\underline r}, {\underline x}_{01}, {\underline b}, Y)$ is
the number density of dipoles of size $x_{01}$ at an impact parameter
${\un b}$ where the momentum fraction of the softest of the two gluons
involved in making up the dipole is greater than $e^{-Y}$. The
quantity $n ({\underline r}, {\underline x}_{01}, {\underline b}, y)$
was defined in \cite{dip} and was shown to satisfy the linear
evolution equation equivalent to BFKL. The solution of that equation
gives \cite{dip}
\be\label{dbn}
\int d^2 B \, n ({\underline r}, {\underline x}_{01}, {\underline B}, Y ) 
\ = \ \int \frac{d \lambda}{\pi i} \, e^{ {\overline \as} \chi (\lambda) 
Y} \, \left( \frac{r}{x_{01}} \right)^{2 \lambda} , 
\ee
where $\chi (\lambda)$ is the eigenvalue of the BFKL kernel
\cite{BFKL} defined as
\be
\chi (\lambda) \, = \, 2 \, \psi (1) - \psi (1 - \lambda) - \psi (\lambda)
\ee
and
\be
{\overline \as} \, = \, \frac{\as N_c}{\pi}.
\ee
Integration in \eq{dbn} runs along a straight line parallel to
imaginary axis to the right of all the singularities of the
integrand. In \eq{levol} $Y$ is the total rapidity interval of the DIS
process while the measured emitted gluon is located at rapidity $y$.

\subsection{Emission of Softer Gluons}

In the previous Subsection we have understood how to include the
evolution of the harder gluons in the virtual photon's wave
function. Here we will address the question of how to include the
evolution of gluons with the light cone component of their momenta
being softer than the measured one. For instance, in \fig{eem} after
being emitted the measured gluon splits the dipole $02$ into two
off-forward dipoles. The dipoles are off-forward because the
transverse coordinate of the measured gluon is not the same in the
amplitude and in the complex conjugate amplitude. Nevertheless there
could still be dipole-like evolution in each of these two
dipoles. More gluons with light cone components of their momenta much
smaller than $k_{3+}$ can be produced and these gluons can interact
with the target as well.

We claim that in order to include the effects of this quantum
evolution (in the leading logarithmic approximation) into \eq{qcincl}
one has to substitute
\be\label{evol}
1 - e^{- {\un x}^2 Q_{0s}^2 ({\un b}) / 4} \, \,  \longrightarrow 
\, \,  N_G ({\un x}, {\un b}, y)
\ee
for all the Glauber exponents in it. The quantity $N_G ({\un x}, {\un
b}, y)$ is the forward amplitude of a {\it gluon} dipole scattering on
the target. This quantity is similar to $N ({\un x}, {\un b}, y)$
except the initial state for $N_G$ consists of a pair of gluons in the
color singlet state instead of a quark-antiquark pair. The
normalization of $N_G$ is analogous to \eq{norm}. In the large-$N_c$
limit an adjoint (gluon) dipole can be decomposed into two fundamental
(quark) dipoles. Therefore the scattering amplitude of a single
adjoint dipole on a target nucleus in the large-$N_c$ approximation is
equivalent to the scattering of two fundamental dipoles on the same
target. One can thus conclude that
\be\label{unitar}
N_G ({\underline x}, {\underline b}, y) \, = \, 2 \, N ({\underline
x}, {\underline b}, y) - N^2 ({\underline x}, {\underline b}, y).
\ee
The evolution equation for $N_G ({\underline x}, {\underline b}, y)$
can be obtained by inverting \eq{unitar} to express $N$ in terms of
$N_G$
\be\label{nng}
N ({\underline x}, {\underline b}, y) \, = \, 1 - \sqrt{1 - N_G^2
({\underline x}, {\underline b}, y)}
\ee
and by substituting \eq{nng} into \eq{eqN}.

One can see right away that without evolution ($y=0$) \eq{evol} turns
into equality (see, for instance, \cite{Mue})
\be\label{ing}
N_G ({\underline x}, {\underline b}, 0) \, = \, 1 - e^{- {\un x}^2
Q_{0s}^2 ({\un b}) / 4}.
\ee
Our goal now is to prove that the substitution of \eq{evol} does
correctly incorporate the effects of subsequent evolution in the gluon
production cross section. As we now know the preceding quantum
evolution happens only at $\tau < 0$ and does not interfere with the
subsequent one. Similarly to Sect. II one should consider four
different cases, corresponding to different emission times of the
measured gluon in the amplitude ($\tau_1$) and in the complex
conjugate amplitude ($\tau_2$). The four cases are: $(1) \,
\tau_1 < 0, \tau_2 < 0; \, \, (2) \, \tau_1 < 0, \tau_2 > 0; \, \, 
(3) \, \tau_1 > 0, \tau_2 < 0; \, \, (4) \, \tau_1 > 0, \tau_2 > 0$.

Let us first consider case $(1)$. In the quasi-classical case of
\eq{disincl} it gave us the exponent in \eq{qc1}. According to \eq{evol} 
we now have to prove that evolution replaces that exponent with $1 -  N_G
\left({\underline z}_1 - {\underline z}_2, \frac{1}{2} ({\underline z}_1 +
{\underline z}_2) , y\right)$. That means that the gluons in the
subsequent evolution connect only to the emitted gluon line mimicking
the scattering amplitude of a gluon dipole of size ${\underline z}_1 -
{\underline z}_2$ on a nucleus.

\begin{figure}
\begin{center}
\epsfxsize=16cm
\leavevmode
\hbox{ \epsffile{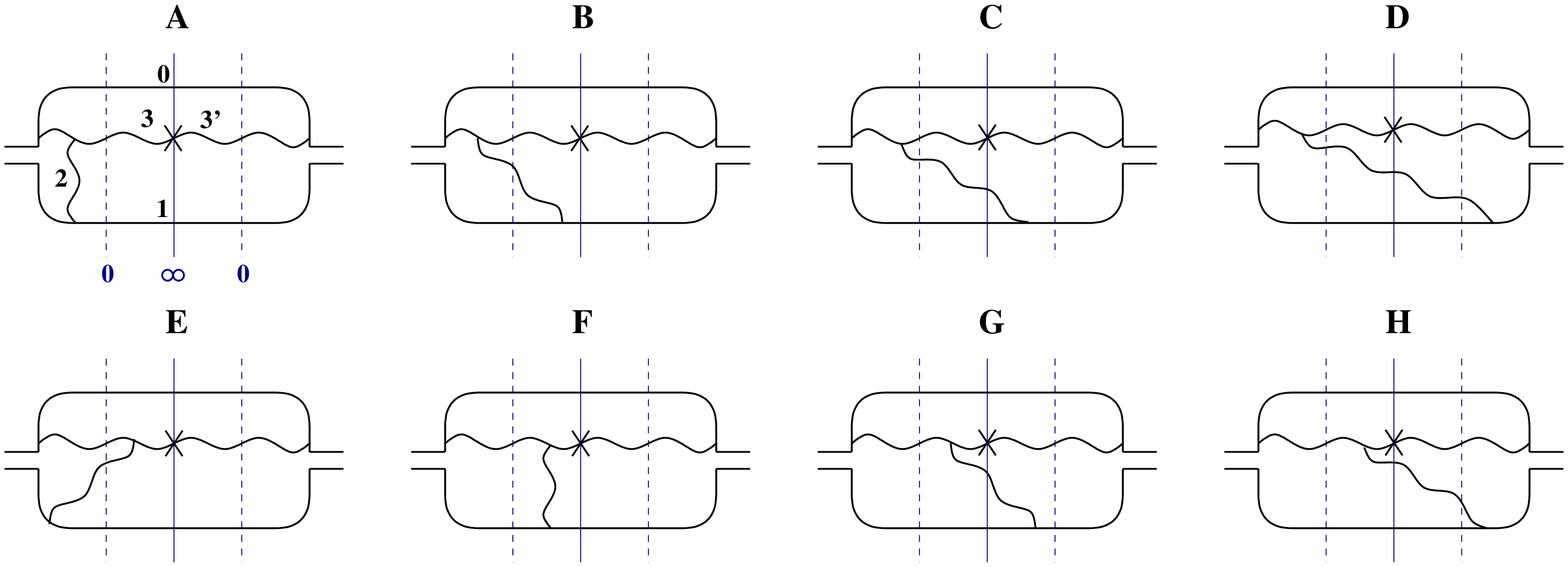}}
\end{center}
\caption{Diagrams including one softer gluon produced after the emission 
of the measured gluon.}
\label{lem}
\end{figure}

To prove that the gluons in the subsequent evolution connect only to
the gluon line of the measured gluon let us again consider a simple
case when we add one more gluon emission to the quasi-classical
diagrams of \fig{qcdis}. Different from the previous Subsection the
extra gluon's light cone momentum should be much softer that the
measured gluon's one. To prove the cancellation of all the diagrams
where the extra soft gluon interacts with the quark lines it is
sufficient to consider graphs presented in \fig{lem}. There after
emitting the measured gluon $\# 3$ we include and extra emission of a
softer gluon $\# 2$ ($k_{2+} \ll k_{3+}$). All the other relevant
diagrams could be obtained by vertical and horizontal reflections of
the graphs in \fig{lem}. In \fig{lem} we kept only the diagrams where
the emission of gluon $\# 2$ is enhanced by a logarithm of energy
leaving out all the subleading ones similar to how it was done in
obtaining \fig{vem}.  The diagrams where the soft gluon ($\# 2$) is
both emitted and absorbed by quark lines are easily canceled by
real-virtual cancellations \cite{dip,cm,kl} and need not be considered
in much detail. In the diagrams A and F in \fig{lem} we imply
summation over both possible time orderings of gluon $\# 2$. As in
\fig{eem} the interaction with the target is not shown explicitly and
is denoted by dashed vertical lines.

Let us consider diagrams A and D in \fig{lem}. Emissions of gluons $\#
2$ and $\# 3$ bring in the same transverse coordinate dependence in
all graphs in \fig{lem}. The only possible difference between A and D
could be in the interactions with the target. In diagram A the Coulomb
gluons could be exchanged only between gluon $\# 3$ and the target,
while it appears that in diagram D the target could interact with
gluon $\# 2$ as well. (Interactions of the target with the quark and
antiquark lines cancel via real-virtual cancellations similar to
\fig{eem}A.) However, since the transverse momentum of gluon $\# 2$ 
is being integrated over the transverse coordinates of that gluon are
equal on both sides of the cut. That way the Coulomb exchanges between
the target and this gluon also cancel through real-virtual
cancellations. Therefore both graphs A and D include the same
interactions with the target giving identical absolute contributions
to the cross section. The only difference is that in graph D the gluon
$\# 2$ is a real gluon while in graph A the gluon $\# 2$ is purely
virtual and is emitted and absorbed at $\tau < 0$. Thus the
contribution of the diagram in \fig{lem}A comes with the negative sign
with respect to the diagram in \fig{lem}D. Therefore they cancel each
other:
\be
A + D \, = \, 0.
\ee
In considering diagrams B and C in \fig{lem} we note that the
interactions with the target are manifestly the same in both of
them. The only difference between graphs B and C is that the gluon has
been moved across the cut in C. Using the cancellation of final state
interaction rules outlined in \cite{cm} we argue that these two
diagrams cancel each other
\be
B + C \, = \, 0.
\ee
Similarly one can show that
\be
F + G \, = \, 0.
\ee
Finally a simple analysis can show that in graphs E and H in \fig{lem}
the interactions with the target involve only gluons and are
identical. In both cases the interactions are equivalent to scattering
of a dipole $32$ and a dipole $3'2$ on the target. Thus the diagrams E
and H give the same absolute contributions. However, their signs are
different since E is virtual while H is real. Therefore they also
cancel
\be
E + H \, = \, 0. 
\ee
We have proved that only interactions with the gluon $\# 3$ survive in
the symmetric case (1). The result could be easily generalized to any
number of softer gluons.

Lastly we have to show that the softer gluons that connect to $3$ and
$3'$ add up to give a scattering amplitude of a gluon dipole $33'$ on
the target nucleus $N_G (3-3')$. To prove this one has to carefully
analyze the diagrams with one extra gluon, similar to \fig{lem} only
with the gluon $\# 2$ connecting only to gluon $\# 3$ ($\# 3'$) and
then generalize the result to any number of soft gluons. This is what
has been checked by the authors. Alternatively we can use the duality
property of the amplitude which was used in \cite{bdmps}. We may just
argue that the amplitude would remain the same if we mirror-reflect
the gluon $\# 3'$ in the complex conjugate amplitude into the
amplitude. Then the gluon production graph would become manifestly
equivalent to the scattering of a $33'$ dipole on the nucleus
justifying the substitution of \eq{evol} in case (1) as desired.

The proofs of substitution of \eq{evol} for cases (2),(3) and (4) is
analogous to the one outlined above. In cases (2) and (3) one first
has to show that emission of a softer gluon is only possible off the
gluon and the quark involved in the appropriate exponents in
\eq{qc2}. That means all interactions with one of the quarks should
cancel. Which one of the quarks becomes the ``spectator'' depends on
the way the gluon $\# 3$ was emitted in the amplitude and in the
complex conjugate amplitude, similar to Sect. II. After that by moving
(reflecting) the remaining (interacting) quark across the cut one can
show that, since in the large-$N_c$ limit two quarks with the same
coordinates are identical to a gluon, the interaction is identical to
a scattering of an adjoint dipole on the target. The adjoint dipole
would be composed of the gluon $\# 3$ and the interacting quark, thus
justifying the substitution of \eq{evol}. In case (4) we explicitly
have two fundamental dipoles $01$ on both sides of the cut developing
the evolution and interacting with the target. Using \eq{unitar} we
can once again prove the substitution of \eq{evol}.

\subsection{Expression for the Inclusive Cross Section}

Now that we have understood above the effects of evolution on the
quasi-classical expression for the inclusive gluon production cross
section we can combine the results to write down the answer including
all evolution effects. The evolution preceding the emission of
measured gluon can be included by the substitution shown in
\eq{levol}. The evolution is linear, similar to what one would obtain 
from AGK cutting rules \cite{glrjet,brjet,agk,gein}. This is not the
first time AGK cutting rules are shown to work for an observable in
dipole evolution. They were also satisfied by the equation for the
diffractive structure function found in \cite{kl}.

The evolution following the emission of the measured gluon is
nonlinear and can be included by making the substitution of
\eq{evol}. It is quite surprising that such a complicated evolution 
effect can be incorporated by such a rather compact rule.

Combining Eqs. (\ref{disincl}) and (\ref{qcincl}) with the
prescriptions of Eqs. (\ref{levol}) and (\ref{evol}) we obtain
\ben
\frac{d \sigma^{\gamma^* A \rightarrow q{\bar q}GX}}{d^2 k \ dy} \ = \ 
\frac{1}{2 \pi^2} \, \int d^2 r  \, d \alpha \,  
\Phi^{\gamma^* \rightarrow q{\bar q}} ({\underline r},\alpha) \ d^2 B \ 
n ({\underline r}, {\underline x}_{01}, {\underline B} - {\underline
b}, Y - y) \, \frac{d^2 x_{01}}{2 \pi x^2_{01}} \, \frac{\as
C_F}{\pi^2} \, \frac{1}{(2 \pi)^2} \, d^2 b
\een
\ben
\times \, d^2 z_1 \, d^2 z_2 \, e^{- i {\underline k} \cdot ({\underline z}_1 -
{\underline z}_2 )} \, \sum_{i,j = 0}^1 (-1)^{i+j} \,
\frac{{\underline z}_1 - {\underline x}_i}{|{\underline z}_1 - 
{\underline x}_i|^2} \cdot \frac{{\underline z}_2 - {\underline
x}_j}{|{\underline z}_2 - {\underline x}_j|^2} \left[ N_G \left({\underline
z}_1 - {\underline x}_j, \frac{1}{2} ({\underline z}_1 + {\underline
x}_j) , y\right) + \right.
\een
\be\label{incl}
+ \left. N_G \left({\underline z}_2 - {\underline x}_i, \frac{1}{2}
({\underline z}_2 + {\underline x}_i), y\right) - N_G
\left({\underline z}_1 - {\underline z}_2, \frac{1}{2} ({\underline
z}_1 + {\underline z}_2) , y\right) - N_G \left({\underline x}_i -
{\underline x}_j, \frac{1}{2} ({\underline x}_i + {\underline x}_j),
y\right)\right],
\ee
where the produced measured gluon has rapidity $y$, the total rapidity
interval in $Y$ and
\be
{\underline b} \, = \, \frac{1}{2} ({\underline x}_0 + {\underline
x}_1).
\ee
\eq{incl} gives us the single inclusive gluon production cross section 
in DIS on a hadron or nucleus including the effects of nonlinear
leading logarithmic evolution and multiple rescatterings. This is the
main result of the paper.

\section{Factorized form of the inclusive cross section}

We now continue by analyzing \eq{incl} in the limit of a very large
target nucleus. In that case the momentum transfer to the nucleus is
cut off by inverse nuclear radius and is very small. We can therefore
take the scattering amplitudes $N_G$ in \eq{incl} at $t=0$, which in
coordinate space is equivalent to neglecting the small compared to the
nuclear radius shifts in impact parameter dependence. Similar to how
it was done in \cite{yurieq,lt} we assume that
\be
N_G ({\un x}, {\un b} \pm {\un y}/2, Y) \, \approx \, N_G ({\un x},
{\un b}, Y),
\ee
where ${\un y}$ represents any of the impact parameter shifts in
\eq{incl}.  Integrating over ${\un z}_1$ or ${\un z}_2$ depending
on the argument of $N_G$ in the term involved we reduce \eq{incl} to
\ben
\frac{d \sigma^{\gamma^* A \rightarrow q{\bar q}GX}}{d^2 k \ dy} \ = \ 
\frac{1}{2 \pi^2} \, \int d^2 r  \, d \alpha \,  
\Phi^{\gamma^* \rightarrow q{\bar q}} ({\underline r},\alpha) \,  \ d^2 B \ 
n ({\underline r}, {\underline x}_{01}, {\underline B} - {\underline
b}, Y - y) \, \frac{d^2 x_{01}}{2 \pi x^2_{01}} \, \frac{\as
C_F}{\pi^2} \, \frac{1}{2 \pi} 
\een
\be\label{fact1}
\times \sum_{i,j = 0}^1 (-1)^{i+j} \, \int d^2 z \, e^{- i {\underline k} \cdot 
{\underline z}} \, \left[ 2 i \, \frac{{\un k}}{{\un k}^2} \cdot
\frac{{\un z} - {\un x}_{ij}}{|{\un z} - {\un x}_{ij}|^2} - 
\ln \frac{1}{|{\un z} - {\un x}_{ij}| \Lambda}  - 2 \pi \, \frac{1}{{\un k}^2} 
\, \delta^2 ({\un z} - {\un x}_{ij}) \right] \, 
\int d^2 b \, N_G ({\un z}, {\un b}, y)
\ee
with ${\un x}_{ij} = {\un x}_i - {\un x}_j$ and $\Lambda$ some
infrared cutoff put in to regulate the integrals
\cite{dip}. Performing summation over $i$ and $j$, employing \eq{dbn} 
and integrating over $x_{01}$ we obtain
\ben
\frac{d \sigma^{\gamma^* A \rightarrow q{\bar q}GX}}{d^2 k \ dy} \ = \ 
\frac{1}{2 \pi^2} \, \int d^2 r  \, d \alpha \,  
\Phi^{\gamma^* \rightarrow q{\bar q}} ({\underline r},\alpha) \, \int 
\frac{d \lambda}{\pi i} \, e^{ {\overline \as} \chi (\lambda) 
(Y-y)} \, \left( \frac{r}{z} \right)^{2 \lambda} \, \frac{\as
C_F}{\pi^2} \, \frac{1}{2 \pi} 
\een
\be\label{fact2}
\times \, d^2 z \, e^{- i {\underline k} \cdot {\underline z}} \, 
\left[ \frac{2 i}{\lambda} \, \frac{{\un k}}{{\un k}^2} \cdot 
\frac{{\un z}}{{\un z}^2} - 
\frac{1}{2 \lambda^2} + \frac{1}{{\un k}^2} \frac{2}{{\un z}^2} \right] \, 
\int d^2 b \, N_G ({\un z}, {\un b}, y). 
\ee
\eq{fact2} can be rewritten as
\ben
\frac{d \sigma^{\gamma^* A \rightarrow q{\bar q}GX}}{d^2 k \ dy} \ = \ 
\frac{1}{2 \pi^2} \, \int d^2 r  \, d \alpha \,  
\Phi^{\gamma^* \rightarrow q{\bar q}} ({\underline r},\alpha) \, \frac{\as
C_F}{\pi^2} \, \frac{1}{2 \pi} \, 
\frac{1}{2 {\un k}^2} \, \int d^2 z \, N_G ({\un z}, y) 
\een
\be\label{fact3}
\times \, \nabla^2_z \, \left[ e^{- i {\underline k} \cdot {\underline z}} \, 
\int \frac{d \lambda}{\pi i} \, e^{ {\overline \as} \chi (\lambda) 
(Y-y)} \, \left( \frac{r}{z} \right)^{2 \lambda} \,
\frac{1}{\lambda^2} \right]
\ee
with
\be
N_G ({\un z}, y) \, = \, \int d^2 b \, N_G ({\un z}, {\un b}, y).
\ee
Integrating by parts yields
\ben
\frac{d \sigma^{\gamma^* A \rightarrow q{\bar q}GX}}{d^2 k \ dy} \ = \ 
\frac{1}{2 \pi^2} \, \int d^2 r  \, d \alpha \,  
\Phi^{\gamma^* \rightarrow q{\bar q}} ({\underline r},\alpha) \, \frac{\as
C_F}{\pi^2} \, \frac{1}{2 \pi} \, 
\frac{2}{{\un k}^2} \, \int d^2 z \, e^{- i {\underline k} 
\cdot {\underline z}} \, \left[ \nabla^2_z \, N_G ({\un z}, y) \right] \, 
\een
\be\label{fact4}
\times \, \frac{1}{\nabla^2_z} \left( \frac{1}{{\un z}^2} \, d^2 B \, 
n ({\un r}, {\un z}, {\un B}, Y-y) \right).
\ee
\eq{fact4} can be rewritten with the help of Lipatov's effective vertex 
\cite{BFKL,brjet}
\be
{\hat L}_k ({\un z}) \, = \, \frac{4 \as N_c}{{\underline k}^2} \,
\stackrel{\textstyle\leftarrow}{\nabla}^2_z \, e^{- i {\underline k} \cdot
{\underline z}} \, \stackrel{\textstyle\rightarrow}{\nabla}^2_z
\ee
as
\ben
\frac{d \sigma^{\gamma^* A \rightarrow q{\bar q}GX}}{d^2 k \ dy} \ = \ 
\frac{1}{\pi (2 \pi)^4} \, \int d^2 r  \, d \alpha \,  
\Phi^{\gamma^* \rightarrow q{\bar q}} ({\underline r},\alpha) 
\een
\be\label{fact}
\times \, \int d^2 z \, N_G ({\un z}, y) \, {\hat L}_k ({\un z}) \, 
\frac{1}{\nabla^4_z} \left( \frac{1}{{\un z}^2} \, d^2 B \, 
n ({\un r}, {\un z}, {\un B}, Y-y) \right).
\ee
\eq{fact} presents our cross section in some sort of $k_T$--factorized form, 
similar to the inclusive cross sections derived in \cite{glrjet}. It
is almost identical with Eq. (14) in \cite{brjet}, which was obtained
by applying AGK rules to BFKL pomeron fan diagrams in DIS case. The
only difference between our \eq{fact} and Eq. (14) in \cite{brjet} is
that the latter uses $N ({\un z}, {\un b}, y)$ (the function $\Phi (y,
r, b)$ in the notation of \cite{brjet}) instead of our $N_G ({\un z},
{\un b}, y)$. That is in order to obtain Eq. (14) of \cite{brjet} from
our \eq{fact} one has to substitute $N_G$ with $N$ in it. In the weak
field limit when the evolution is linear both $N$ and $N_G$ obey the
same BFKL evolution equation, though with different initial conditions
given by Eqs. (\ref{gla}) and (\ref{ing}) correspondingly. In the
saturation region the evolutions equations for the two quantities in
question become different. $N ({\un z}, {\un b}, y)$ obeys \eq{eqN},
while the equation for $N_G ({\un z}, {\un b}, y)$ can be obtained by
substituting \eq{nng} into \eq{eqN}. The resulting equation would
involve square roots and would be quite different from
\eq{eqN}.  However the high energy asymptotics of the two objects is
still the same: both $N ({\un z}, {\un b}, y)$ and $N_G ({\un z}, {\un
b}, y)$ asymptotically approach $1$ at high energies, when $y
\rightarrow \infty$.

The form of the cross section in \eq{fact} suggests that in a certain
gauge or in some gauge invariant way it could be written in a
factorized form involving two unintegrated gluon distributions merged
by effective Lipatov vertex. The usual form of the factorized
inclusive cross section is \cite{claa}
\be\label{kt1}
\frac{d \sigma}{d^2 k \ dy} \ = \ \frac{2 \as}{C_F \, {\un k}^2} \, 
\int d^2 q \, \frac{f_1 (\xi_1, {\un q}^2) \, 
f_2 (\xi_2, |{\un k} - {\un q}|^2)}{{\un q}^2 \, ({\un k} - {\un q})^2}
\ee
with $\xi_1$ and $\xi_2$ the values of Bjorken $x$ variable for the
gluons in each of the colliding particles. $f_1 (\xi_1, {\un q}^2)$
and $f_2 (\xi_2, |{\un k} - {\un q}|^2)$ are the unintegrated gluon
distributions
\be\label{unint}
f_i (\xi_i, {\un q}^2) \, = \, \frac{d \, xG_i (\xi_i, {\un q}^2)}{d \ln
{\un q}^2}.
\ee
\eq{kt1} can be written in the coordinate space as
\be\label{kt2}
\frac{d \sigma}{d^2 k \ dy} \ = \ \frac{2 \pi^2}{N_c C_F} \, \int d^2 z 
\, \phi_1 (\xi_1, {\un z}) \, {\hat L}_k ({\un z}) \, \phi_2 (\xi_2, {\un z})
\ee
with
\be\label{four}
 \phi_i (\xi_i, {\un z}) \, = \, \int \frac{d^2 q}{(2 \pi)^2} \, e^{i {\un
 z} \cdot {\un q}} \, \frac{f_i (\xi_i, {\un q}^2)}{{\un q}^4}.
\ee
Comparing \eq{kt2} with the gluon production cross section in
dipole-nucleus scattering which follows from \eq{fact}
\be\label{fact5}
\frac{d \sigma^{q{\bar q} A \rightarrow q{\bar q}GX}}{d^2 k \ dy} \ = \ 
\frac{1}{(2 \pi)^3} \, \int d^2 z \, N_G ({\un z}, y) \, {\hat L}_k ({\un z}) \, 
\frac{1}{\nabla^4_z} \left( \frac{1}{{\un z}^2} \, d^2 B \, 
n ({\un r}, {\un z}, {\un B}, Y-y) \right)
\ee
and assuming that the latter was generated through the same
factorization mechanism we identify
\be\label{id1}
N_G ({\un z}, y = \ln 1/\xi) \, \Leftrightarrow \, \frac{(2
\pi)^{4} \, \as}{N_c} \, \phi (\xi, {\un z}).
\ee
Employing Eqs. (\ref{unint}) and (\ref{four}) we may rewrite \eq{id1}
as
\be\label{id2}
\nabla^2_z N_G ({\un z}, y = \ln 1/\xi) \, = \frac{(2
\pi)^2 \, \as}{N_c} \, \int d^2 q \, e^{i {\un
 z} \cdot {\un q}} \, \frac{d \, xG (\xi, {\un q}^2)}{d {\un q}^2},
\ee
which corresponds to the definition of gluon distribution used in
\cite{braun,brjet}. However \eq{id2} is not satisfied in the
quasi-classical limit of no evolution. There $N_G ({\un z}, 0)$ is
given by \eq{ing} integrated over impact parameter
\be\label{ngcl}
N_G ({\un z}, y=0) \, = \, S_\perp \left(1 - e^{ - {\underline z}^2 \,
Q_{0s}^2 / 4 } \right),
\ee
while the gluon distribution including all multiple rescatterings have
been calculated in \cite{KM} to give
\be\label{xgcl}
\int d^2 q \, e^{i {\un
 z} \cdot {\un q}} \, \frac{d \, xG_{cl} (\xi, {\un q}^2)}{d {\un
 q}^2} \, = \, \frac{2}{\pi} \, \int d^2 b \, \mbox{Tr} \left< {\un
 A}^{WW} ({\un 0}) \cdot {\un A}^{WW} ({\un z}) \right> \, = \,
 \frac{2 S_\perp C_F}{\pi^2 \as \, {\underline z}^2} \, \left(1 - e^{ -
 {\underline z}^2 \, Q_{0s}^2 / 4 } \right),
\ee
where ${\un A}^{WW} ({\un z})$ is the non-Abelian
Weisz\"{a}cker-Williams field of a nucleus \cite{yuri} and we assumed
that the nucleus is cylindrical with the cross sectional area $S_\perp
= \pi R^2$. Expanding both \eq{ngcl} and \eq{xgcl} to the lowest order
in $Q_{0s}^2$ corresponding to two gluon exchange we can see that
\eq{id2} can be easily satisfied. However full Eqs. (\ref{ngcl}) and 
(\ref{xgcl}) when inserted into \eq{id2} do not satisfy it. Therefore
\eq{id2} seems to work for the leading twist two gluon exchange
approximation but appears to fail once we include multiple
rescatterings in it.

The failure of \eq{id2} in the quasi--classical limit makes the physical
meaning of factorization of \eq{fact5} quite obscure. The factorized
form of \eq{fact5} implies convolution of two unintegrated gluon
distributions with Lipatov vertex as shown in \eq{kt1}. It appears,
however, that we can not identify one of the convoluted functions
($N_G$) with the unintegrated gluon distribution in disagreement with
the factorization hypothesis of \eq{kt1}. This observation by itself
would be quite natural indicating that higher twists in the form of
multiple rescatterings modify the relationship between $N_G$ and the
unintegrated gluon distribution. On the other hand the fact that
$k_T$--factorization is still preserved seems somewhat unexpected. The
reason why we are able to write down the inclusive cross section of
\eq{incl} in the factorized form of \eq{fact} needs to be better clarified 
which will be done elsewhere.

\section{Properties of the Inclusive Cross Section}

It was shown in \cite{lt} that asymptotic solutions to the nonlinear
evolution equation bear most of quantitative features of the general
one and are very convenient for discussion of various high energy
processes.  To study asymptotic behavior of the inclusive gluon
production it is convenient to integrate over directions of vector $z$
in \eq{fact2} and use Fourier transform of the gluon scattering
amplitude $\tilde N_G(l,y)$ defined as
\begin{equation}\label{tilda}
N_G(\un z,y)\, =\, \un z^2\, \int_0^\infty\, dl\, l\, J_0(lz)
\tilde N_G(l,y)
\end{equation}
with $l = |{\un l}|$, obtaining  
\ben
\frac{d \sigma^{\gamma^* A \rightarrow q{\bar q}GX}}{d^2 k \ dy} \ = \
\frac{1}{2\pi^2}\int d^2r\, d\alpha\,\Phi^{\gamma^*\rightarrow
  q\bar q}(\underline r,\alpha) \int\, \frac{d\lambda}{2\pi i}
e^{\bar\alpha_s\chi(\lambda)(Y-y)}\,\frac{\alpha_s C_F}{\pi^2}
\int_0^\infty\frac{dz}{z}\,\left(\frac{r}{z}\right)^{2\lambda}
\een
\begin{equation}\label{factor}
\times\frac{z^2}{k^2}\int_0^\infty\, dl\, l\, J_0(l z)\,\tilde N_G(l,y)
\left[\frac{2kz}{\lambda}J_1(kz)-
\frac{k^2z^2}{2\lambda^2}J_0(kz)+2J_0(kz)\right].
\end{equation}

\begin{figure}
\begin{center}
\leavevmode
\epsfxsize=6cm
\hbox{\epsffile{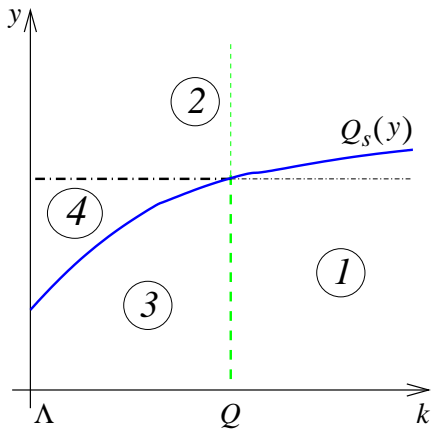}}
\end{center}
\caption{Various asymptotic kinematical regions discussed in the text.
Solid line is the saturation scale $Q_s(y)$. Bold lines are the
boundaries of different regions.}
\label{fig:diagram}
\end{figure}

There are four interesting asymptotic regions of \eq{factor}. They are
shown in \fig{fig:diagram}. In the first and second regions the
produced gluon momentum $k$ is the largest external momentum scale of
the process. The first region in \fig{fig:diagram} corresponds to the
cases when $k \gg Q_s (y) \gg Q$ and $k \gg Q \gg Q_s (y)$, while in
the second region $Q_s (y) \gg k \gg Q$ and $Q_s (y) \gg Q \gg k$. We
can find asymptotic behavior of the inclusive cross section by
expanding function $\chi(\lambda)$ near its simple pole at $\lambda
=0$ and evaluating the Mellin transform in $\lambda$ in the saddle
point approximation. This corresponds to the double logarithmic
approximation to the evolution equation which means summation of
$\alpha^n\ln^n k \ (Y-y)^n\sim 1$ terms. The third and fourth regions in
\fig{fig:diagram} depict the cases when $Q \gg k \gg Q_s (y)$ and $Q
\gg Q_s (y) \gg k$ correspondingly.  In the third and fourth regions
asymptotic behavior of the inclusive cross section can be found by
repeating this procedure near the simple pole of $\chi(\lambda)$ at
$\lambda =1$ which again corresponds to summation of $\alpha^n\ln^n k \
(Y-y)^n\sim 1$ terms. These are two asymptotic double logarithmic
regions relevant to the linear evolution before the gluon has been
emitted. Generally, after its emission the evolution is nonlinear. So,
there is a kinematical domain where the parton density is large
(regions 2 and 4 in \fig{fig:diagram}). In this domain a color dipole
evolves into two dipoles, one of which is of the size $1/Q_s(y)$.  And
there is a kinematical domain where the parton density is small
(regions 1 and 3) and the evolution is still linear.

Integration over $z$ in \eq{factor} yields a combination of
generalized hypergeometric functions which can be expanded near
$\lambda=0$ and $\lambda=1$ to give
\begin{equation}\label{zint}
\int_0^\infty\frac{dz}{z}\, \left(\frac{r}{z}\right)^{2\lambda}
z^2J_0(lz)\left[\frac{2kz}{\lambda}J_1(kz)-
\frac{k^2z^2}{2\lambda^2}J_0(kz)+2J_0(kz)\right]=
\left\{ \begin{array}{c}
 \frac{4(k r)^{2\lambda}}{k^2 \lambda},\; \lambda\rightarrow 0, \, k \gg l,
\\ \\ \\ \frac{4 k^2 (l r)^{2\lambda}}{l^4 \lambda},\; \lambda\rightarrow 0, \, 
k \ll l, \\ \\ \\ \frac{(k r)^{2\lambda}}{k^2(1-\lambda)},\;
\lambda\rightarrow 1, \, k \gg l, \\ \\ \\ 
\frac{(l r)^{2\lambda}}{l^2(1-\lambda)} ,\; \lambda\rightarrow 1, \, 
k \ll l. \end{array} \right.
\end{equation} 
Each of the limits in \eq{zint} corresponds to the appropriate region
in \fig{fig:diagram}. It is now straightforward to evaluate the Mellin
transform in the saddle point approximation, which corresponds to the
double logarithmic approximation of the scattering amplitude. Using
\eq{zint} in \eq{factor} we obtain in the region with $k \gg l$ and 
$\lambda = 0$
\begin{equation}\label{leqz}
\frac{d\sigma^{\gamma^* A \rightarrow q{\bar q}GX}}{d^2k\, dy}=
\frac{8}{(2\pi)^3}\frac{\bar\alpha_s}{\pi}\frac{\sqrt{\pi}}{\un k^4}
\int\, d^2r\,d\alpha\,  \Phi^{\gamma^* \rightarrow q{\bar q}}(\un r,\alpha)
\frac{e^{2\sqrt{\bar\alpha_s(Y-y)\ln(k^2r^2)}}}
{ \left(\bar\alpha_s\ln(k^2r^2)(Y-y)\right)^{1/4} }\,
\int_0^k\, dl\, l\, \tilde N_G(l,y),
\end{equation}
We used the well-known expansion
$\chi(\lambda)=\lambda^{-1}+\mathcal{O} (\lambda)$. Scattering
amplitude $N_G(l,y)$ must be normalized to give the correct Fourier
transformed  initial condition \eq{ngcl}. We have (neglecting
logarithmic dependence of $Q_s$ on $l$)
\begin{equation} 
\tilde N_G(l,0)=\frac{Q_0^2}{l^2}\int\frac{d\lambda}{2\pi i}
\left(\frac{l}{Q_0}\right)^{2\lambda}2^{-2\lambda-1}
\left(\frac{Q_0}{Q_{0s}}\right)^{2\lambda-2}\frac{\Gamma(1-\lambda)}{1-\lambda}
=\frac{1}{2}\Gamma \left(0, \frac{l^2}{Q_s^2}\right),
\end{equation}
where $\Gamma (0,z)$ is the incomplete gamma function of zeroth order.
In the region 1, where $l\gg Q_s$, evolution is linear and the
scattering amplitude is small. In the linear region
\begin{equation}\label{nglin}
N_G(l,y)=\frac{Q_{0s}^{2} \pi R^2}{2l^2}\int 
\frac{d\nu}{2\pi i}\left(\frac{l}{Q_{0}}
\right)^{2\nu} e^{\frac{\bar\alpha_s}{\nu}y}=
\frac{Q_{0s}^{2} \pi R^2}{2\pi\,l^2}
\frac{\sqrt{\pi}\bar\alpha_s^{1/4}\, y^{1/4}}
{\ln^{3/4}(l^2/Q_{0}^2)} e^{2\sqrt{\bar\alpha_s
y\ln(l^2/Q_0^2)}},\quad \ln(l/Q_0)\gg
\bar\alpha_s y,
\end{equation} 
where we expanded $\chi(\nu)$ near $\nu=0$ and assumed the nucleus to
be a cylinder with the cross sectional area $S_\perp = \pi R^2$. $Q_0$
is an initial scale for linear evolution. \eq{nglin} gives gluon
scattering amplitude in double logarithmic approximation with the
expansion parameter $\bar\alpha_s\, y \ln(l/Q_0)\sim 1$. Upon
substitution of \eq{nglin} into \eq{leqz} and integration over $l$ one
arrives at the asymptotic expression for the inclusive cross section
in the region 1
\begin{equation}\label{reg1}
 \left. \frac{d\sigma^{\gamma^* A \rightarrow q{\bar q}GX}}{d^2k\, dy}
 \right|_{\mbox{region}\, 1}=
\frac{\pi R^2}{(2\pi)^4}\frac{\bar\alpha_s\,Q_{0s}^{2}}{2\un k^4}
\int\, d^2r\,d\alpha\,  \Phi^{\gamma^* \rightarrow q{\bar q}}(\un r,\alpha)
\frac{e^{ 2\sqrt{\bar\alpha_s(Y-y)\ln(k^2r^2)} +
2\sqrt{\bar\alpha_s y\ln(k^2/Q_{0s}^2)}}} {\bar\alpha_s^{1/2} \, [y
(Y - y)]^{1/4} (\ln k^2/Q_{0s}^2)^{1/4}\ln^{1/4}(k^2r^2)}.
\end{equation}

In the region 2 with $l\ll Q_s$ saturation sets in and the total
scattering amplitude $N(\un x,y)$ is close to unity. Thus, using
\eq{unitar} we conclude that $N_G(\un x,y)$ is close to unity as
well. In the momentum representation this is equivalent to the
following asymptotic behavior in the saturation regime \cite{lt}
\begin{equation}\label{ngsat}
\tilde N_G(l,y)=\ln(Q_s(y)/l), \quad l \ll Q_s (y).
\end{equation} 
Using \eq{ngsat} together with the second line of \eq{zint} in
\eq{leqz} yields
\begin{equation}\label{reg2}
\left. \frac{d\sigma^{\gamma^* A \rightarrow q{\bar q}GX}}{d^2k\, dy} 
\right|_{\mbox{region}\, 2}=
\frac{16 \pi R^2}{(2\pi)^4}\sqrt{\pi} \frac{\bar\alpha_s}{\un k^2}\ln 
\frac{Q_s^2 (y)}{k^2}
\int\, d^2r\,d\alpha\,  \Phi^{\gamma^* \rightarrow q{\bar q}}(\un r,\alpha)
\frac{e^{2\sqrt{\bar\alpha_s (Y-y)\ln(k^2r^2)}}}
{\left(\bar\alpha_s\, (Y-y)\ln(k^2r^2)\right)^{1/4}}.
\end{equation}

Consider now $\lambda=1$ pole of function $\chi(\lambda)$ in
\eq{factor}. In the vicinity of this point
$\chi(\lambda)=(1-\lambda)^{-1}+
\mathcal{O}$$(1-\lambda)$.
Substituting the third line of \eq{zint} into \eq{zint} and repeating
the above procedure we end up with following asymptotics in the third
kinematical region. In the region 3 ($Q \gg k\gg Q_s$)
\begin{equation}\label{reg3}
\left. \frac{d\sigma^{\gamma^* A \rightarrow q{\bar q}GX}}{d^2k\, dy} 
\right|_{\mbox{region}\, 3} \hspace{-1.5mm}=
\frac{\pi R^2}{(2\pi)^4}\frac{\bar\alpha_s\,Q_{0s}^{2}}{\un k^2}
\int d^2r\,d\alpha\,  \Phi^{\gamma^* \rightarrow q{\bar q}}(\un r,\alpha)
\, r^2\, 
\frac{e^{ 2\sqrt{\bar\alpha_s(Y-y)\ln(1/k^2r^2)} +
2\sqrt{\bar\alpha_s y\ln(k^2/Q_{0s}^2)}}} {\bar\alpha_s^{1/2} [y
(Y-y)]^{1/4}\left(\ln( k^2/Q_{0s}^2)\ln(1/k^2r^2)\right)^{1/4}}.
\end{equation}
Inserting the last line of \eq{zint} into \eq{factor} and using
\eq{ngsat} together with the expression for $N_G$ generated by linear
evolution gives for the region 4 ($k\ll Q_s \ll Q$)
\begin{equation}\label{reg4}
\left. \frac{d\sigma^{\gamma^* A \rightarrow q{\bar q}GX}}{d^2k\, dy}
\right|_{\mbox{region}\, 4}=
\frac{4\pi R^2}{(2\pi)^4}\frac{\bar\alpha_s\,Q_s^2(y)}{\un k^2}\sqrt{\pi}
\int\, d^2r\,d\alpha\,  \Phi^{\gamma^* \rightarrow q{\bar q}}(\un r,\alpha)
\, r^2\, \frac{e^{2\sqrt{\bar\alpha_s\, (Y-y)\ln(1/Q_s^2 (y) r^2)}}}
{\left(\bar\alpha_s\, (Y-y)\ln(1/Q_s^2 (y) r^2)\right)^{1/4}},
\end{equation}
where the integration over $l$ has been done with logarithmic
accuracy.

Note that the spectrum of produced gluons falls off as $\sim k^{-4}$
in region~1, which is a well-known result of the perturbation
theory. One factor of $k^{-2}$ arises from the Lipatov's effective
vertex, while another one is given by the perturbative behavior of the
scattering amplitude $\nabla^2 N_G$ which scales as $\sim k^{-2}$ at
large $k_\perp$.  When the typical momentum $l$ of the late evolution
is less than the saturation scale $Q_s(y)$, the anomalous dimension of
the gluon structure function is close to unity. Thus the scattering
amplitude depends only logarithmically on momentum. As a result, in
the region~2 the gluon spectrum softens to $k^{-2}$ behavior. In the
region~3 the late evolution is linear like in region~1. However an
additional factor of $k^2$ stems from different double logarithmic
asymptotics at $\lambda=1$. It is remarkable that unlike the regions~1
and 2, the spectrum in region~4 drops in the same manner as in the
region~3 despite the fact that the late evolution is nonlinear. This
happens because the typical value of $l$ in the rightmost integral in
\eq{leqz} is $l\sim Q_s$. Due to this fact \eq{reg4} is enhanced by
a factor of $Q_s^2$ rather than $k^2$. Of course, one could not expect
that the spectrum will depend on produced gluon's momentum only
logarithmically in the presence of a hard projectile (virtual photon)
which wave function has not reached saturation yet.

\begin{figure}
\begin{center}
\leavevmode
\epsfxsize=13cm
\hbox{\epsffile{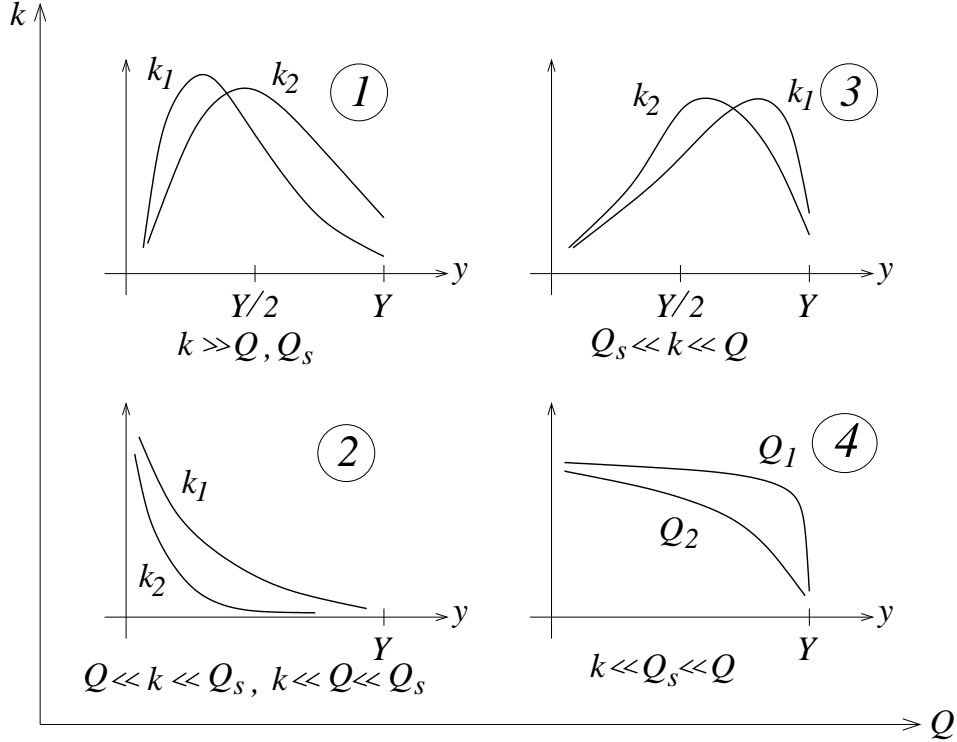}}
\end{center}
\caption{Qualitative energy dependence of the inclusive gluon 
production cross section. All lines are scaled by different numerical
factors to underline important features of the spectrum. In all plots
we show the curves for two different values of $k$ and/or $Q$
($k_1<k_2$, $Q_1<Q_2$).}
\label{ydep}
\end{figure}

Derived asymptotic formulae \eq{reg1}, \eq{reg2}, \eq{reg3} and
\eq{reg4} make it possible to understand some important features of the
spectrum energy dependence. Equating the $y$ derivative of the
expression in the exponent of \eq{reg1} (or \eq{reg3}) to zero one
finds the position of the inclusive spectrum maximum $y_0$ at some
fixed $k$ in the double logarithmic approximation in which the
pre-exponential factor is a slowly varying function. It reads
\begin{equation}\label{max1}
y_0=Y\frac{1}{1\pm\frac{\ln^2(kr)}{\ln^2(k/Q_{0s})}}
\stackrel{k\gg Q, Q_{0s}}{\longrightarrow}\frac{1}{2}Y,
\end{equation}
where $+$ and $-$ correspond to the regions 1 and 3.  By analogy we
find that in the region~2 spectrum is a monotonically decreasing
function of $y$ and exhibits no maximum. In the region~4 the spectrum
has no maximum in $y$ as well. The following equation defines a
surface in the space of parameters where spectrum is constant:
\begin{equation}\label{max2}
\left(4\bar\alpha_s Y+\ln^2\frac{1}{Q_s^2(0)r^2}\right)^2=
16\bar\alpha_s Y \ln\frac{1}{Q_s^2(0)r^2}.
\end{equation}
Although \eq{max2} is beyond the validity of the double logarithmic
approximation employed in this section, it shows the general feature
of the spectrum $y$-dependence in region~4. Namely, when $Q\sim Q_s$
the spectrum is a step-like function.  In \fig{ydep} we show a
qualitative picture of the energy behavior of the inclusive cross
section given by \eq{incl} in different kinematical regions.
Note, that all four plots in \fig{ydep} show different qualitative
behavior of the inclusive cross section as the rapidity varies. By
plotting the experimental data on the energy dependence of the
spectrum one can distinguish between different kinematical regions
which may be useful for evaluation of the saturation scale.  However,
we realize that our calculations in the asymptotic regions presented
in this Section are approximate and an exact numerical analysis of the
cross section in \eq{fact} has to be performed to enable us to
describe the data with it.

\section{Conclusions}

We have constructed an expression for single inclusive gluon
production cross section in DIS (\eq{incl}) including multiple pomeron
exchanges in the form of nonlinear evolution (\eq{eqN}). The cross
section may be used to describe (mini)jet production in DIS.  For the
case of a large target nucleus the resulting production cross section
can be written in $k_T$--factorized form of \eq{fact}. The transverse
momentum spectrum given by the cross section of \eq{incl} reproduces
the usual perturbative behavior $\sim 1/{\un k}^4$ in the large
$k_\perp$ limit (see Eqs. (\ref{reg1}) and (\ref{reg2})). In the small
transverse momentum region the spectrum softens to $\sim 1/{\un k}^2$
(see Eqs. (\ref{reg3}) and (\ref{reg4})). The cross section still
exhibits some residual infrared divergence. This is due to the fact
that we are scattering a point-like probe (virtual photon) on the
target. The target nucleus' wave function has reached saturation and
this is why the $k_\perp$ dependence softens in the infrared. However,
for the cross section to be infrared safe (up to logarithms), similar
to \cite{yuaa}, we need the wave functions of both colliding particles
to reach saturation. In \cite{yuaa} that was reached for the case of
nucleus-nucleus scattering where both nuclear wave functions were in
the saturation region. In our case the wave function of the
quark-antiquark pair has not reached saturation yet. The onset of
saturation in the $\gamma^*$ wave function would again be due to
multiple pomeron exchanges.  Unlike the case of a nucleus with
nucleons there is no extra non-perturbative color charges in the
original incoming $\gamma^*$ wave function to facilitate
saturation. All the color charges have to be generated
perturbatively. Therefore the multiple pomeron exchanges in the
$\gamma^*$ wave function would take the form of pomeron loops
\cite{dip,robi}. Summation of pomeron loops is a separate problem and
is beyond the scope of this paper. The effects of pomeron loops can be
safely neglected in the energy range considered here.

The result of \eq{incl} could be generalized to the case of
proton-nucleus scattering (pA). If one models proton as a dipole made
out of diquark and a quark the inclusive gluon production cross
section can be obtained by appropriately modifying the linear
evolution term (\eq{levol}) and changing the $\gamma^*$ wave function
into the proton's wave function. In a more generic case of a proton
consisting of $N_c$ valence quarks the generalization is probably a
bit more involved, though the preceding evolution still remains linear
and the gluon production mechanism would not be modified.

We have shown that the evolution in the nuclear wave function (at
least in the large-$N_c$ limit) can be included via a simple
substitution of the Glauber-Mueller amplitude by the dipole-evolved
amplitude as presented in \eq{evol}. We may therefore try to apply
this principle to nucleus-nucleus collisions (AA). In \cite{yuaa} it
was argued that the multiplicity of produced gluons in a central
nuclear collision is likely to be given by the following formula:
\ben
\frac{d N^{AA}_{cl}}{d^2 k \, d^2 b \, dy} \ = \ \frac{2 \, C_F}{\as \pi^2} \, 
\left\{  -  \int \frac{d^2 z}{(2 \pi)^2} \,  e^{i {\underline k} \cdot 
{\underline z}} \, \frac{1}{{\underline z}^2} \, \left(1 - e^{ -
{\underline z}^2 \, Q_{0s1}^2 (b) / 4 } \right) \, \left(1 - e^{ -
{\underline z}^2 \, Q_{0s2}^2 (b) / 4 } \right) + \right.
\een
\ben
+ \left. \int \frac{d^2 x \ d^2
y}{(2 \pi)^3} \, e^{i {\underline k} \cdot ({\underline x} -
{\underline y})} \,  \frac{{\underline x}}{{\underline x}^2}
\cdot \frac{{\underline y}}{{\underline y}^2} \,  \left[ \frac{1}{{\underline x}^2 
\, \ln \frac{1}{|{\underline x}| \mu}} \, \left(1 - e^{ - {\underline x}^2 
\, Q_{0s1}^2 (b) / 4 } \right) \, \left(1 - e^{ - {\underline x}^2
\, Q_{0s2}^2 (b) / 4 } \right) + \right. \right.
\een
\be\label{aasol}
+ \left. \left.  \frac{1}{{\underline y}^2 
\, \ln \frac{1}{|{\underline y}| \mu}} \, \left(1 - e^{ - {\underline y}^2 
\, Q_{0s1}^2 (b) / 4 } \right) \, \left(1 - e^{ - {\underline y}^2
\, Q_{0s2}^2 (b) / 4 } \right) \right] \right\}.
\ee
$Q_{0s1}^2 (b)$ and $Q_{0s2}^2 (b)$ are the saturation scales in each
of nuclei taken at the same impact parameter since the collisions
considered are central.  Inspired by the success of the substitution
(\ref{evol}) in incorporating the quantum evolution corrections in DIS
we may conjecture the following ansatz for the multiplicity
distribution of the produced gluons in AA including nonlinear
evolution in the wave functions of both nuclei
\ben
\frac{d N^{AA}}{d^2 k \, d^2 b \, dy} \ = \ \frac{2 \, C_F}{\as \pi^2} \, 
\left\{  -  \int \frac{d^2 z}{(2 \pi)^2} \,  e^{i {\underline k} \cdot 
{\underline z}} \, \frac{1}{{\underline z}^2} \, N_{G1} ({\un z}, {\un
b}, y) \, N_{G2} ({\un z}, {\un b}, Y-y) +
\right.
\een
\ben
+ \left. \int \frac{d^2 x \ d^2
y}{(2 \pi)^3} \, e^{i {\underline k} \cdot ({\underline x} -
{\underline y})} \,  \frac{{\underline x}}{{\underline x}^2}
\cdot \frac{{\underline y}}{{\underline y}^2} \,  \left[ \frac{1}{{\underline x}^2 
\, \ln \frac{1}{|{\underline x}| \mu}} \, N_{G1} ({\un x}, {\un
b}, y) \, N_{G2} ({\un x}, {\un b}, Y-y) + \right. \right.
\een
\be\label{aaevol}
+ \left. \left.  \frac{1}{{\underline y}^2 
\, \ln \frac{1}{|{\underline y}| \mu}} \, \, N_{G1} ({\un y}, {\un
b}, y) \, N_{G2} ({\un y}, {\un b}, Y-y) \, \right] \right\},
\ee
where $Y$ is the total rapidity interval and $N_{G1}$ and $N_{G2}$ are
adjoint dipole forward scattering amplitudes in the first and second
nucleus correspondingly. At the moment we can not prove \eq{aaevol}
and leave it as an ansatz inspired by the properties of the inclusive
cross sections in the saturation region studied above. We may argue
that the final state gluon mergers which are not included in
\eq{aasol} are not likely to give logarithms of energy and therefore
should not contribute to the quantum evolution.  They might be
neglected compared to the evolution effects included in
\eq{aaevol}. \eq{aaevol}, together with \eq{eqN}, may be used to 
describe the emerging RHIC data similar to how it was done in
\cite{dg}.

\section*{Acknowledgments}

The authors would like to thank Mikhail Braun, Genya Levin, and Al
Mueller for stimulating and informative discussions.

The work of Yu. K. was supported in part by the U.S. Department of
Energy under Grant No. DE-FG03-97ER41014 and by the BSF grant $\#$
9800276 with Israeli Science Foundation, founded by the Israeli
Academy of Science and Humanities. The work of K. T. was sponsored in
part by the U.S. Department of Energy under Grant
No. DE-FG03-00ER41132.


\begin{thebibliography}{99}

\bibitem{glrmq} L. V. Gribov, E. M. Levin, and M. G. Ryskin,
  Phys. Rep. {\bf 100}, 1 (1983); A.H. Mueller,
J.-W. Qiu, Nucl. Phys. {\bf B268}, 427 (1986).

\bibitem{LR}
E.M.  Levin and M.G Ryskin, Nucl. Phys. {\bf B304}, 805 (1988);
Sov. J.  Nucl. Phys. {\bf 45}, 150 (1987); {\bf 41}, 300 (1985).

\bibitem{Mue}
A.H. Mueller, Nucl. Phys. {\bf B335}, 115 (1990). 

\bibitem {mv} 
L.\ McLerran and R.\ Venugopalan, Phys.\ Rev.\ D {\bf 49}, 2233
(1994); {\bf 49}, 3352 (1994); {\bf 50}, 2225 (1994).

\bibitem{yuri} 
Yu.V.\ Kovchegov, Phys.\ Rev.\ D {\bf 54}, 5463 (1996); {\bf 55}, 5445
(1997).

\bibitem {jklw} 
J.\ Jalilian-Marian, A.\ Kovner, L.\ McLerran, and 
H.\ Weigert, Phys. Rev. D {\bf 55},5414 (1997). 

\bibitem{yurieq} 
Yu. V. Kovchegov, Phys. Rev. D {\bf 60}, 034008 (1999); D {\bf 61},
074018 (2000).

\bibitem{dip}
A.H.\ Mueller, Nucl.\ Phys.\ {\bf B415}, 373 (1994); A.H.\ Mueller and
B. Patel, Nucl.\ Phys.\ {\bf B425}, 471 (1994); A.H.\ Mueller, Nucl.\
Phys.\ {\bf B437}, 107 (1995). 

\bibitem{cm}
Z.\ Chen, A.H.\ Mueller, Nucl.\ Phys.\ {\bf B451}, 579 (1995).

\bibitem{bal}
I. I. Balitsky, Report No. hep-ph/9706411; Nucl. Phys. {\bf B463}, 99
(1996); Phys. Rev. D {\bf 60}, 014020 (1999).

\bibitem{JKLW} J. Jalilian-Marian, A. Kovner, A. Leonidov, and H.
  Weigert, Nucl. \ Phys. {\bf B 504}, 415 (1997); Phys.\ Rev. {\bf D
  59} 014014 (1999); Phys. Rev. {\bf D59}, 034007 (1999); J.
  Jalilian-Marian, A. Kovner, and H.  Weigert, Phys.\ Rev. {\bf D 59}
  014015 (1999); A. Kovner, J. G. Milhano, H. Weigert, Phys. Rev. D
  {\bf 62}, 114005 (2000); H. Weigert, hep-ph/0004044; E. Iancu,
  A. Leonidov, L. McLerran, Nucl. Phys. {\bf A692}, 583 (2001);
  Phys. Lett. {\bf B510}, 133 (2001); E. Iancu and L. McLerran,
  Phys. Lett. {\bf B510}, 145 (2001); E. Ferreiro, E. Iancu,
  A. Leonidov, L. McLerran, hep-ph/0109115.

\bibitem{lt}
E. Levin and K. Tuchin, Nucl. Phys. {\bf B573}, 833 (2000);
Nucl. Phys. {\bf A691}, 779 (2001); Nucl. Phys. {\bf A693}, 787
(2001).

\bibitem{braun}
M. A. Braun, Eur. Phys. J. C {\bf 16}, 337 (2000); hep-ph/0010041;
hep-ph/0101070.

\bibitem{agl} 
A.L.Ayala, M.B.Gay Ducati, E.M.Levin, Nucl. Phys. {\bf B493}, 305
(1997); Nucl. Phys. {\bf B511}, 355 (1998); Eur. Phys. J. {\bf C8},
115 (1999).

\bibitem{lm}
C.S. Lam, G. Mahlon, Phys. Rev. D{\bf 61}, 014005 (2000);
Phys. Rev. D{\bf 62}, 114023 (2000); Phys. Rev. D{\bf 64}, 016004
(2001).

\bibitem{lub}
E. Levin and M. Lublinsky, hep-ph/0104108; M. Lublinsky,E. Gotsman, E. 
Levin and U. Maor, hep-ph/0102321. 

\bibitem{gbm}
K. Golec-Biernat, L. Motyka, A.M. Stasto, hep-ph/0110325. 

\bibitem{musat}
A. H. Mueller, Nucl. Phys. {\bf B572}, 227 (2000); Nucl. Phys. {\bf
B558}, 285 (1999); hep-ph/0111244. 

\bibitem{KM}
Yu. V. Kovchegov, A.H. Mueller, Nucl. Phys. {\bf B529}, 451 (1998). 

\bibitem{BFKL}
E.A. Kuraev, L.N. Lipatov and V.S. Fadin, {\em Sov. Phys. JETP} {\bf
45}, 199 (1977); Ya.Ya. Balitsky and L.N. Lipatov, {\em Sov. J. Nucl.
Phys.} {\bf 28}, 22 (1978).

\bibitem{yuincl}
Yu. V. Kovchegov, Phys. Rev. D {\bf 64}, 114016 (2001).  

\bibitem{bb}
I.I. Balitsky, A.V. Belitsky, hep-ph/0110158. 

\bibitem{diffqc}
A. Hebecker, Nucl.Phys. {\bf B505}, 349 (1997); W. Buchm\"{u}ller,
M. F. McDermott, and A. Hebecker, Phys. Lett. B {\bf 410}, 304 (1997);
W. Buchm\"{u}ller, T. Gehrmann, and A. Hebecker, Nucl.Phys. {\bf
B537}, 477 (1999); Yu. V. Kovchegov and L. McLerran, Phys. Rev. D {\bf
60}, 054025 (1999); Erratum-ibid. D {\bf 62}, 019901 (2000).

\bibitem{kl}
Yu. V. Kovchegov, E. Levin, Nucl. Phys. {\bf B577}, 221 (2000). 

\bibitem{claa} 
A.\ Kovner, L.\ McLerran, and H.\ Weigert, Phys.\ Rev.\ D {\bf 52},
6231 (1995); {\bf 52}, 3809 (1995); M. Gyulassy, L. McLerran,
Phys. Rev. C {\bf 56}, 2219 (1997); Yu.V.\ Kovchegov, D. H. Rischke,
Phys. Rev. C {\bf 56}, 1084 (1997); X. Guo, Phys. Rev, D {\bf 59},
094017 (1999).

\bibitem{gb}
J.F.\ Gunion and G.\ Bertsch, Phys.\ Rev.\ D {\bf 25}, 746 (1982).

\bibitem{kop}
B. Z. Kopeliovich, A. V. Tarasov, A. Schafer, Phys. Rev. C {\bf 59},
1609 (1999); B. Z. Kopeliovich, A. Schafer, A. V.  Tarasov,
Phys. Rev. D {\bf 62}, 054022 (2000); B. Kopeliovich, A. Tarasov,
J. Hufner, hep-ph/0104256.

\bibitem{md}
A. Dumitru, L. McLerran, hep-ph/0105268; A. Dumitru,
J. Jalilian-Marian, hep-ph/0111357.

\bibitem{kw}
A. Kovner, U. A. Wiedemann, Phys. Rev. D {\bf 64}, 114002 (2001).

\bibitem{kv} A. Krasnitz, R. Venugopalan, Report No. hep-ph/0007108; 
Phys. Rev. Lett. {\bf 84}, 4309 (2000); Nucl. Phys. {\bf B557}, 237
(1999); A. Krasnitz, Y. Nara, R. Venugopalan, Phys. Rev. Lett. {\bf
87}, 192302 (2001). 

\bibitem{yuaa}
Yu. V. Kovchegov, Nucl. Phys. {\bf A692}, 557 (2001); Nucl. Phys. {\bf
A698}, 619c (2002).

\bibitem{ab}
N. Armesto, M.A. Braun, hep-ph/0107114. 

\bibitem{glrjet}
L. V. Gribov, E. M. Levin, and M. G. Ryskin, Phys. Lett. {\bf B100},
173 (1981).

\bibitem{brjet}
M. A. Braun, Phys. Lett. {\bf B483}, 105 (2000).  

\bibitem{agk}
V. A. Abramovsky, V. N. Gribov, O. V. Kancheli,
Sov. J. Nucl. Phys. {\bf 18}, 308 (1974).

\bibitem{gein}
E. M. Levin, hep-ph/9710546. 

\bibitem{brod}
S. J. Brodsky, B.-Q. Ma, Phys. Lett. {\bf B381}, 317 (1996);
S. J. Brodsky, P. Hoyer, C. Peterson, N. Sakai, Phys. Lett. {\bf B93},
451 (1980); S. J. Brodsky, C. Peterson, N. Sakai, Phys. Rev. D {\bf
23}, 2745 (1981).

\bibitem{brod1}
G. P. Lepage, S. J. Brodsky, Phys. Rev. D {\bf 22}, 2157 (1980);
S. J. Brodsky, R. Roskies, R. Suaya, Phys. Rev. D {\bf8}, 4574 (1973);
P. P. Srivastava, S. J. Brodsky, Phys. Rev. D {\bf 64}, 045006 (2001).

\bibitem{robi}
H. Navelet, R. Peschanski, C. Royon, S. Wallon, Phys. Lett. {\bf
B385}, 357 (1996); H. Navelet, R. Peschanski, Phys. Rev. Lett. {\bf
82}, 1370 (1999); A. Bialas, H. Navelet, R. Peschanski,
Nucl. Phys. {\bf B593}, 438 (2001).

\bibitem{bdmps}
R. Baier, Yuri L. Dokshitzer, A.H. Mueller, S. Peigne, D. Schiff,
Nucl. Phys. {\bf B483}, 291 (1997).

\bibitem{dg}
D. Kharzeev, E. Levin, nucl-th/0108006; D. Kharzeev, E. Levin,
M. Nardi, hep-ph/0111315.

\end{thebibliography}
\end{document}